\documentclass[sigconf]{acmart}

\usepackage{multirow}
\usepackage{makecell}
\usepackage{graphicx}
\usepackage{xcolor}
\usepackage{hyperref}

\AtBeginDocument{%
  }

\setcopyright{acmlicensed}

\copyrightyear{2025}
\acmYear{2025}
\setcopyright{cc}
\setcctype{by}
\acmConference[CHI '25]{CHI Conference on Human Factors in Computing Systems}{April 26-May 1, 2025}{Yokohama, Japan}
\acmBooktitle{CHI Conference on Human Factors in Computing Systems (CHI '25), April 26-May 1, 2025, Yokohama, Japan}\acmDOI{10.1145/3706598.3713417}
\acmISBN{979-8-4007-1394-1/25/04}

\newcommand{\revised}[1]{{#1}}
\newcommand{\camready}[1]{{#1}}
\begin{document}

\title{VideoDiff: Human-AI Video Co-Creation with Alternatives}


\author{Mina Huh}
\authornote{This work was done during the author's Adobe research internship}
\affiliation{
  \institution{The University of Texas at Austin}
  \country{}}
\email{minahuh@cs.utexas.edu}

\author{Dingzeyu Li}
\affiliation{
  \institution{Adobe Research}
  \country{}}
\email{dinli@adobe.com}

\author{Kim Pimmel}
\affiliation{
  \institution{Adobe Research}
  \country{}}
\email{kipimmel@adobe.com}

\author{Hijung Valentina Shin}
\affiliation{
  \institution{Adobe Research}
  \country{}}
\email{vshin@adobe.com}

\author{Amy Pavel}
\affiliation{
  \institution{The University of Texas at Austin}
  \country{}}
\email{apavel@cs.utexas.edu}

\author{Mira Dontcheva}
\affiliation{
  \institution{Adobe Research}
  \country{}}
\email{mirad@adobe.com}


\begin{abstract}
To make an engaging video, people sequence interesting moments and add visuals such as B-rolls or text. While video editing requires time and effort, AI has recently shown strong potential to make editing easier through suggestions and automation. A key strength of generative models is their ability to quickly generate multiple variations, but when provided with many alternatives, creators struggle to compare them to find the best fit. We propose VideoDiff, an AI video editing tool designed for editing with alternatives. With VideoDiff, creators can generate and review multiple AI recommendations for each editing process: creating a rough cut, inserting B-rolls, and adding text effects. VideoDiff simplifies comparisons by aligning videos and highlighting differences through timelines, transcripts, and video previews. Creators have the flexibility to regenerate and refine AI suggestions as they compare alternatives. Our study participants (N=12) could easily compare and customize alternatives, creating more satisfying results.
\end{abstract}




\begin{teaserfigure}
  \centering
  \includegraphics[width=\textwidth]{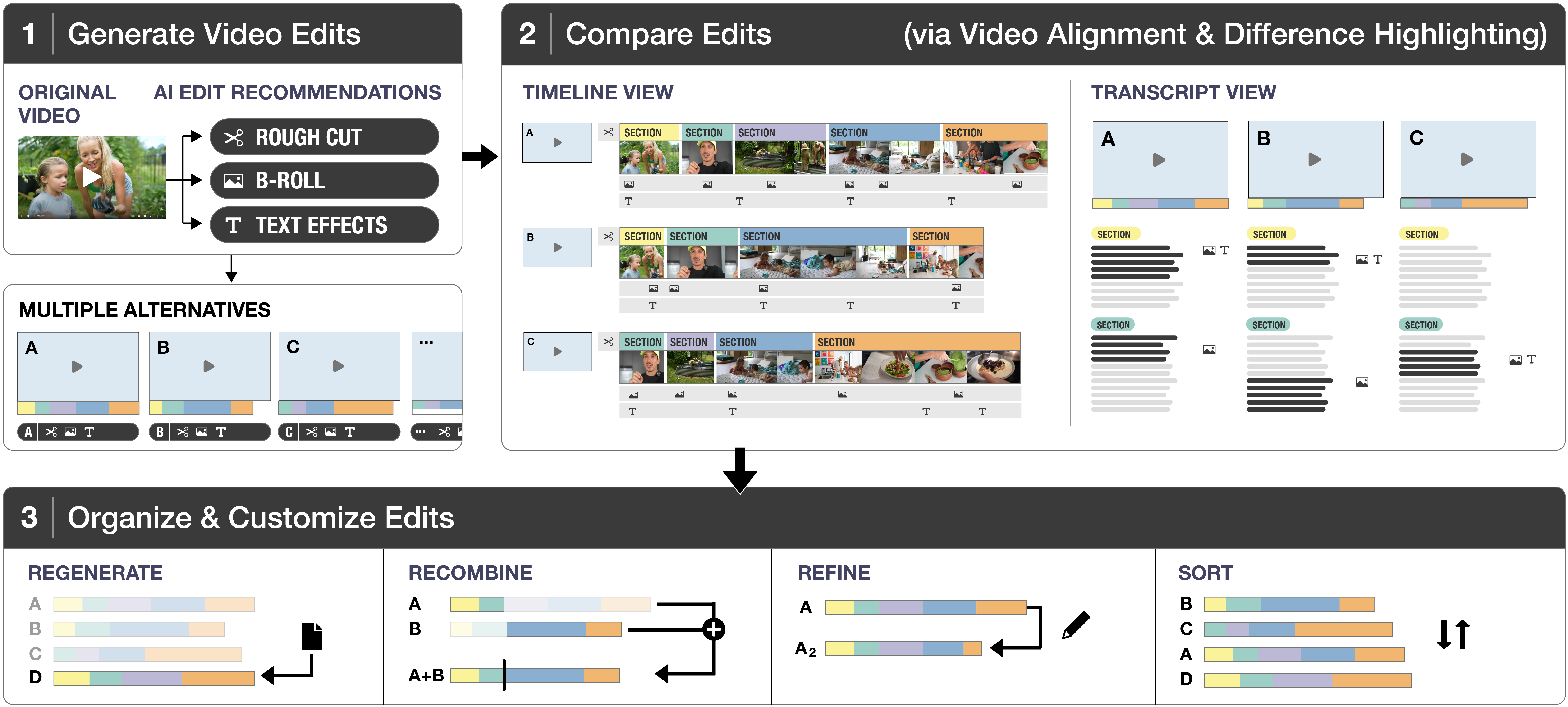}
  \caption{VideoDiff is a Human-AI co-creative system that supports video creators to explore multiple variations. 1) VideoDiff generates diverse AI recommendations for making rough cuts, inserting B-rolls, and adding text effects. 2) VideoDiff supports easy comparison by aligning videos and highlighting differences using timeline and transcript views. 3) Users can narrow down options by sorting and further customize them by refining and regenerating AI suggestions.}\label{fig:teaser}
\end{teaserfigure}

\newcommand{\ipstart}[1]{\vspace{1mm}\noindent{\textbf{\textit{#1.}}}}
\newcommand\smallverb[1]{\texttt{\small #1}}

\newcommand{\sysname}{VideoDiff}
\maketitle

\section{Introduction}
As video has become a mainstream form of communication and storytelling, an increasing number of end users create and share videos. YouTube, which is considered the most popular video-focused social network, now has approximately 64 million creators worldwide~\cite{YTstats}.
Yet creating compelling videos is a complex and time consuming task. Creators need to find key moments~\cite{wang2024podreels} and remove irrelevant and repetitive content~\cite{fried2019text, huh2023avscript}. They also spend time making the video more engaging with visual effects, B-roll~\cite{huber2019b}, text~\cite{xie2023wakey}, and music~\cite{rubin2014generating}. 
Recent advances in video understanding and generative models have shown great potential for assisting video editing.  Prior research has shown that AI tools can speed up multiple stages of video authoring including script writing~\cite{mirowski2023co}, storyboarding~\cite{wang2024reelframer}, asembling clips into rough cuts~\cite{chi2013democut, truong2016quickcut}, identifying low-quality footage~\cite{huh2023avscript, fried2019text}, and adding B-roll~\cite{huber2019b}. 
Recent AI video products such as OpusClip~\cite{opus}, CapCut~\cite{capcut}, and Vizard~\cite{vizard} further streamline video editing by automatically making cuts, and adding transition effects and subtitles. 

One powerful new capability enabled by generative AI models is the quick generation of multiple variations. This allows creators to explore many alternative stories or B-roll placements simultaneously, potentially resulting in better final videos~\cite{dow2010parallel, suh2024luminate}. \revised{While most existing video editing tools are designed to handle only one video version at a time, recent AI tools such as OpusClip~\cite{opus} and Capcut~\cite{capcut} generate multiple variations of edited videos to offer diverse options to users. Despite the benefits of exploring alternatives in creative tasks, there are potential new burdens to users:} 1) comparing the variations~\cite{huh2023genassist, gero2024supporting} and 2) managing them over time~\cite{reza2023abscribe, suh2024luminate}.
While prior work explored sensemaking and comparison of multiple AI generations in text~\cite{gero2024supporting, suh2023sensecape, reza2023abscribe}, images~\cite{almeda2024prompting, huh2023genassist}, and designs~\cite{swearngin2020scout, matejka2018dream}, comparing multiple videos presents unique challenges due to the temporal nature of video. In this work, we explore this emerging approach to video editing centered on working with multiple variations. 
To understand the opportunities and challenges of authoring videos with multiple alternatives, we conducted a formative study where 8 professional video creators were tasked with comparing multiple edited videos of the same source content. \camready{
Creators in our study mentioned that comparison is a common practice throughout their current video editing process as they consider alternative narratives, visual assets, or lengths of videos.
They also highlighted that having alternatives helps them reflect on their preferences and further plan editing directions, but it was time-consuming to manually create multiple versions. As AI speeds up the video creation process, we
envision that future video editing tools will more commonly provide users with multiple variations.}

However, reviewing alternatives can require a lot of time and effort because creators must watch all the videos to compare them. Unlike static images or text, which can be quickly scanned for differences, videos unfold over time, requiring creators to watch, pause, and replay to detect variations in stories, transitions, and visual effects. Based on these findings, we derive 6 design goals for video editing tools centered on working with alternatives. 
A tool to support creators in exploring video alternatives should minimize redundant watching (D1) and enable quick skimming of differences (D2) across different stages of the editing process (D3) in the most appropriate modality (D4). Such a tool should also help creators easily organize and customize variations to support them in finding a version with fewer errors (D5) and best suits their needs (D6). 
\revised{We introduce ~\textit{VideoDiff}, an AI-powered video editing tool designed to enhance the creative potential of working with alternatives while minimizing the burden of comparison.} VideoDiff generates diverse AI recommendations for each video editing task such as making rough cuts, inserting B-rolls, and adding text effects (Figure~\ref{fig:teaser}.1). 
For each task, creators can quickly skim the differences between variations using VideoDiff which aligns the variations and highlights the differences using timelines, transcripts, and video previews (Figure~\ref{fig:teaser}.2). 
Additionally, creators can sort variations to narrow down the options and further customize them by refining and regenerating AI suggestions (Figure~\ref{fig:teaser}.3). 
In this work, we focus specifically on working with narrative videos such as lifestyle vlogs, podcasts, how-to videos, and travel videos. 
Their descriptive dialogue makes them well suited to the current strengths of text-based large language models, which are good at extractive and abstractive summarization.  Additionally, the source content for these types of videos is either one long video (\textit{e.g.,} presentation or an event) or multiple sources that can be sequenced in time because they capture chronological activities, such as a how-to demonstration or travel videos. This simplifies our visualization design to show how different versions relate to one another. 




We evaluated VideoDiff in a within-subjects study with 12 video creators who compared VideoDiff with a baseline interface designed to encompass features of existing AI video editing tools (\textit{e.g.,} CapCut, OpusClip). Participants rated VideoDiff as more useful for quickly understanding the differences of multiple videos which helped them create and consider more diverse variations. Participants also expressed higher satisfaction with the final videos created with VideoDiff.
The contributions of our paper are as follows:
\begin{itemize}
    \item We identify the challenges of comparing multiple variations in video editing and derive design goals for systems to support video creation with alternatives.
    \item We propose \textit{VideoDiff}, an AI-powered video editing tool that supports video creation with alternatives by enabling efficient generation, comparison, organization, and customization of variations.
    \item Our user study findings showcase the advantages and challenges of using VideoDiff.
\end{itemize}

\section{Related work}

\subsection{Human-AI video co-creation}

Traditional video editing tools such as Premiere Pro~\cite{PremierePro} or iMovie~\cite{iMovie} require users to manually inspect the video, plan for what edits to make, and execute them by interacting with the complex UI in video editing tools. With the power of AI to understand the content and quality of videos, new forms of video editing systems have been introduced that speed up and automate the video creation process~\cite{invideo, opus, capcut, Descript}. 
Earlier systems supported users to structure narrated videos into multiple segments to help them quickly skim and select appropriate parts~\cite{truong2016quickcut, truong2019tool, wang2019write, leake2017computational, huh2023avscript}.
These systems label clips with metadata such as speaker and topic to let users navigate and search~\cite{truong2019tool, huh2023avscript}, or recommend video segments that are semantically relevant to the script~\cite{truong2016quickcut, wang2019write, xia2020crosscast, leake2020generating}.
Beyond organization of clips, researchers have also proposed tools that recommend edits based on the video content and quality. B-Script~\cite{huber2019b} analyzes the video transcript and recommends possible B-roll positions and clips. 
AVscript~\cite{huh2023avscript} detects low-quality footage (\textit{e.g.,} blur, bad lighting) to cut out, and commercial tools such as Descript~\cite{Descript} and Vimeo~\cite{Vimeo} identify filler words and pauses for quick removal. 
Beyond post-editing, prior research has also proposed automatic video creation from various media including text documents~\cite{chi2022synthesis}, webpage~\cite{chi2020automatic}, markdown~\cite{chi2021automatic}, news articles~\cite{wang2024reelframer}, and music~\cite{liu2023generative}. 

More recent tools proposed LLM-powered video editing tools that are capable of understanding users' editing intent in natural language and automatically executing them. ExpressEdit~\cite{tilekbay2024expressedit} allows users to edit videos by typing edit instructions and sketching on the video frame.
LAVE~\cite{wang2024lave} provides an LLM-agent that assists users in planning and performing edit actions while also enabling users to manually refine agent actions. \
To support the task of repurposing long videos into short and compelling videos, PodReels~\cite{wang2024podreels} and ROPE~\cite{wang2022record} explore the use of LLMs to create podcast summaries. 

Existing Human-AI video co-creation tools already create multiple variations for users to better explore the design space. For instance, ReelFramer~\cite{wang2024reelframer} suggests multiple narrative framing to let users explore and choose video topics, and ROPE~\cite{wang2022record} generates summaries of variable lengths. However, the new task of how users review, refine, and select among these variations of videos remains unexplored. 
In this era of AI-driven video creation, we investigate user challenges in reviewing and comparing AI-edited videos.

\subsection{Video summarization, skimming, and browsing}

Understanding and browsing video content is an important first step in video editing. Traditional video players commonly use a ~\textit{seek bar} which indicates the progress of the video and allows viewers to navigate to different parts of the video. To address the difficulty of previewing upcoming scenes, many video players on the web including YouTube provide picture-in-picture previews of thumbnails when viewers hover over the seek bar and video chapters that segment video into multiple sections and add context to each portion for quick skimming.

There is a large body of work that summarizes a video as images by selecting the important keyframes~\cite{truong2016quickcut, chang2021rubyslippers, barnes2010video, boreczky2000interactive, jin2017elasticplay}. 
Video editing tools such as Premiere Pro and iMovie utilize ~\textit{filmstrip}, a visual representation of a video's frames displayed in a linear sequence. This format allows editors to get a quick overview of the entire video at a glance and identify parts for cutting or inserting effects. To enable seamless zoom over filmstrip, Video Tapestries eliminated borders between frames for spatial continuity~\cite{barnes2010video}. 
Other works used grid-view of keyframes to summarize the videos~\cite{boreczky2000interactive, jin2017elasticplay}. Boreczkly et al. proposed a comic book presentation for videos that displayed keyframes with different sizes based on their importance~\cite{boreczky2000interactive} and ElasticPlay visualizes an interactive summary based on dynamic time budget using a grid of thumbnails~\cite{jin2017elasticplay}.

Researchers have also explored video browsing interfaces that leverage video transcripts. To support the navigation of informational videos, these systems highlight keywords in the transcript~\cite{chang2021rubyslippers}, or provide scaffolding steps~\cite{yang2022softvideo, fraser2020temporal} and concepts~\cite{liu2018conceptscape}.
Building upon prior research on video summarization and browsing techniques, our work explores visual representations to support efficient review of multiple edited videos. 


\subsection{Comparison and sensemaking}
\camready{
Comparison is crucial in creative tasks, helping users understand the design space and refine their decisions. Variation theory~\cite{marton2014necessary} suggests that learning occurs through experiencing contrasts. Thus, exposing creators to alternatives can help them understand model affordances and make more informed decisions.~\cite{gebreegziabher2024supporting}}

To visualize comparison, 3 main techniques are used: juxtaposition, superposition, and explicit encoding~\cite{gleicher2011visual}.
\textit{Juxtaposition} places distinct visual elements side-by-side to highlight similarities and differences. \textit{Superposition} overlays visuals within a common frame and  \textit{Explicit encoding} reveals the predefined relationship between visualizations. 
Prior work utilized these techniques for video comparison by building sequential and parallel video players~\cite{baker2024interaction, tharatipyakul2018towards}, and highlighting the subtle differences with color overlay~\cite{balakrishnan2015video, baker2024interaction}. 
To support the search of educational videos, VSedu supports video comparison based on topic relevance~\cite{benedetto2024visual}. Video Lens facilitated the search in a large collection of sports videos via rapid playback and event-based search~\cite{matejka2014video}. This line of work extends information visualization techniques to support the comparison of videos from multiple sources. Yet, they do not support video comparison in the editing context, where the comparison involves multiple variations of edited videos from the same source footage. 

\revised{As AI continues to speed up and automate content generation, it is becoming more common to work with multiple variations in parallel in creative tasks including writing~\cite{reza2023abscribe, suh2024luminate, buschek2021impact}, image generation~\cite{dalle3, midjourney, huh2023genassist}, video generation~\cite{sora}, and video editing~\cite{opus, capcut}. While facilitating divergent thinking and decision-making, this new workflow of working with variations demands an additional stage of comparison. This has led researchers to explore ways to compare and understand AI-generated content.} 
Tools have provided multiple levels of abstraction to support users in managing complex information generated by AI~\cite{suh2023sensecape, liu2024selenite}.
\camready{To facilitate collaboration with AI in creative tasks, researchers explored sensemaking and comparison of writing variations~\cite{reza2023abscribe, suh2024luminate, koch2014varifocalreader}, images~\cite{huh2023genassist, almeda2024prompting, koch2020imagesense, brade2023promptify}, slides~\cite{drucker2006comparing}, storyboards~\cite{benedetto2024visual}, and designs~\cite{matejka2018dream}. However, video presents unique challenges due to the temporal nature and multi-modal complexity. We contribute an interface that supports sensemaking of multiple variations of videos and provide empirical insights into how creators utilize variations when co-creating videos with AI.}




\section{Formative study}
To better understand how video creators work with alternatives and the support they need for comparing and managing them, we conducted a formative study with 8 professional video editors. The formative study consisted of semi-structured interviews and a comparison task with 3 videos.

\subsection{Method}
\ipstart{Participants}
We recruited 8 professional video creators using Upwork\footnote{\url{https://www.upwork.com/}} (P1-P8, \S{A} Table~\ref{tab:participants}). We compensated the participants with their self-set hourly rate (\$28, SD=5.3) for a 1-hour remote study conducted via Zoom. Participants had an average of 10.5 years of experience (SD=3.0) and created a wide variety of videos including commercials, interviews, documentaries, vlogs, and short-form videos.

\ipstart{Procedure}
We started the study with demographic and background questions about the participants' video editing experiences and current approaches for reviewing and comparing videos.
In the comparison task, participants reviewed and compared 3 videos edited from the same source footage. We pre-uploaded videos to YouTube so that users could use a familiar interface that offered speed controls, hover thumbnails, and the video transcript. 
To become familiar with the original version, participants first skimmed the source footage. Then, the participants reviewed the 3 edited versions for 30 minutes. 
They were allowed to take notes to help them remember any details. At the end of the task, participants were asked to summarize the differences between the three videos and choose one they preferred, explaining their choice. We conducted a post-task interview to understand their strategies for comparing videos and any challenges they encountered.

\ipstart{Materials}
For the comparison task, we used 2 different source videos (F1: cooking tutorial~\cite{cooking_video}, F2: TED talk~\cite{ted_talk_video}).  We chose these videos aiming for source footage that was largely unedited and varied in duration (F1: 1hr 11min, F2: 12 min) and visual information (F1: featuring many objects, F2: primarily talking head).
We used 2 approaches to create multiple edited versions of the source video -- 1) with AI and 2) with experts. This allowed us to see how users work with variations created today with the current capabilities of AI video tools and in the future when AI capabilities improve.
For AI edits, we created 3 versions of the summary for each source video using an LLM to extract subsets of the video. 
Using the video transcript, we instructed GPT-4 to generate 3 summaries on the video topic (F1: How to make Eggs Benedict, F2: Talk highlights). 
We also hired 6 professional video editors to manually create 3 edited versions of F1 and F2~\footnote{See Supplementary Materials for examples.}.
The video editors made a variety of editing decisions including trimming content to enhance focus, adding subtitles, B-roll footage, and motion graphics, incorporating text overlays and background music, and making audio enhancements. Each participant was given 3 edited versions from the same source video created with the same approach either with AI or manually. For F1, LLM edited videos were 152 seconds (SD=24) and expert edited videos were 451 seconds (SD=152). For F2, LLM edited videos were 283 seconds (SD=110) and expert edited videos were 323 seconds (SD=21).

\ipstart{Analysis}
To understand how participants explore alternatives, we analyzed the study recordings and participants' notes. We transcribed the interviews and participants' spontaneous comments during the task and grouped the transcript according to 1) current practices, 2) comparison strategies, and 3) opportunities and challenges of video editing via comparison. 

\subsection{Findings}
\ipstart{Current Practices}
Participants reported that decision-making via comparison is a common process throughout the video editing process. They mentioned having to review and select which clips to include from the footage provided by their clients (P1, P3-4, P7-P8), decide on the appropriate B-roll images (P2, P4), and experiment with different logo styles or placements (P5-P6).

Participants often had to create multiple versions of videos upon clients' requests (P1, P3-P8). 
This was because clients often found it challenging to imagine the final look of the videos from the editors' descriptions alone. P5 and P6 who created ad videos for marketing companies created two versions for companies' A/B testing to see which version is more effective. 
Clients often requested videos that have different lengths, storylines, intro versions, caption styles, or music styles. P3 and P4 mentioned that clients often liked different aspects of multiple versions so they had to combine the versions in further editing. 
While creating multiple versions helped clients narrow down and make decisions, it required a lot of time and effort from video editors. P5 said ~\textit{``I know they (clients) want to see three completely different videos, but I try not to because it takes forever. I only tweak small things that involve minor changes and can be done quickly.''}
As a workaround, participants created variations and got intermediate feedback in each stage (\textit{e.g.,} sharing multiple variations of scripts before moving on to the storyboard) so that they could minimize the amount of rework required later in the process (P1, P5, P8). Alternatively, P4 and P6 created multiple versions of the introductory part of a video instead of the full videos as introductions can capture the essence of what the rest of the videos will look like.

\ipstart{Comparison Strategies - Watching Videos}
In the comparison task, participants watched the 3 versions of the edited videos to understand their differences. 5 participants watched each video from beginning to end before moving on to the next video (P1-P4, P6) while 3 participants switched to play other videos (P5, P7-P8). P7 first watched the introductions of all three videos because it was easier to remember and compare shorter segments and the introductions often captured the holistic style of the video. P5 and P8 switched to other versions of edited videos when they were not satisfied with the current version's edits (\textit{e.g.,} text effects) to check if other videos offered better alternatives. 
While reviewing edited videos, P5 revisited the original unedited video to check which part of the source video was missing. 4 participants re-watched the same video multiple times to identify parts they missed in the first review attempt (P3, P5-P6, P8). 

\ipstart{Comparison Strategies - Taking Notes}
7 participants took notes while reviewing edited videos to better recall the differences (P2-P8). They took notes on the content covered or omitted (\textit{e.g., yolk separation step is not included}), errors (\textit{e.g., mid-sentence jumps, flash frames}), what they liked or disliked about the edits applied (\textit{e.g., Nice intro music, Too many text effects}), how they would have edited differently (\textit{e.g., I would have emphasized the end results by showing it longer}), and how the video compared to other versions (\textit{e.g., This is better at showing the sequence of events compared to the first one}). 
Participants reviewing videos edited with LLM noted errors in the edits, such as sentence cut-offs and jump cuts.
P1 did not take any notes because pausing and taking notes would make the reviewing process much longer.

\ipstart{Challenges}
All participants mentioned the comparison process to be very time-consuming as they had to watch multiple videos to identify the differences. As a workaround, 5 participants sped up the video and skipped around the videos to find the edit points (P3-4, P6-P8).
Although the transcripts of the edited videos were provided, participants noted that simply reviewing the transcript or thumbnail previews on the timeline cannot fully replace the watching experience. P1 explained~\textit{``Simply reading the transcript isn't enough because the transcript might seem fine, but the visuals might have been messed up.''} P3 who checked for edit errors in the video noted that errors such as sentence cut-offs or jump cuts cannot be spotted using the transcript or by scrubbing through the thumbnails; she had to watch the entire video to identify them.

Comparing edited videos was also mentally demanding as participants had to compare multiple aspects of videos at the same time (\textit{e.g.,} storyline, visual effects, B-rolls, subtitles, music). 
As a result, participants often reviewed each version multiple times to focus on reviewing a single aspect at a time (P3, P5-P6, P8). P5 described ~\textit{``I cannot focus on multiple aspects at once while reviewing. When I’m checking the colors, I only look at colors.''}
When comparing the content, it was challenging to switch between multiple videos as they extracted different parts of the original video. For instance, P3 wanted to check if all videos included a specific cooking step but found it difficult as the step appeared in different timestamps across the videos. 

Participants mentioned that it requires a lot of effort to manually identify edits made throughout the video, take notes to remember details, and understand both high-level and low-level differences among videos from their notes. In the post-task interview, P5 mentioned that he wanted AI to describe the high-level theme or goal of each edited version to help him better understand what he was reviewing and what to expect. P6 wanted to have a list of all the edits made in each video so that he would not miss subtle edits (\textit{e.g.,} color correction, small text effects) and could quickly jump to those specific edit points for preview.



\subsection{Design Guidelines}
Video editors in our formative study currently make decisions by comparison throughout the video editing process and mentioned the benefits of having multiple variations of edited videos. We envision that in near future, Human-AI video co-creation tools will more commonly provide users with multiple variations. 
Our formative study reveals design guidelines (D1-D6) to better support video creation with alternatives. Our work aims to address D1-D4 and D6 while leaving D5 as a future work.

\begin{itemize}
    \item[\textbf{D1.}] \textbf{Minimize redundant watching by aligning variations} 
    Participants found it difficult to repeatedly watch overlapping sections of edited videos to identify edit points and differences. They wanted to compare edited variations with the original video and across alternatives but found it challenging with mismatched timelines. Aligning variations can simplify comparisons and reduce redundant viewing. 
    
    \item[\textbf{D2.}] \textbf{Support quick skimming by highlighting differences}
    Prior work explored comparison in text~\cite{reza2023abscribe, gero2024supporting} or images~\cite{huh2023genassist, almeda2024prompting}, but the temporal aspect of videos makes it challenging to skim them for comparison. This can be time-consuming and tedious as video lengths increase or as more videos are compared. A system that describes or visualizes the differences among edited videos can help users quickly make decisions.
    \item[\textbf{D3.}] \textbf{Enable independent comparison for different editing stages} Compared to traditional video editing where editors go through an interactive process of making sequential edits and reviewing, recent AI products such as CapCut or OpusClip speed up this process and generate fully edited videos. The resulting videos can be mentally demanding to compare, as users have to consider multiple aspects (\textit{e.g.,} rough cuts, subtitle styles) together. Enabling users to compare aspects independently can reduce the cognitive load.
     \item[\textbf{D4.}] \textbf{Support comparison via multiple modalities} Relying on a single medium, like transcripts or thumbnails, is insufficient for comparing video variations. Edits in visuals or timing can be unnoticed when only text is reviewed. Supporting comparison through multiple modalities—such as transcripts, visual thumbnails, and audio—enables users to identify various types of differences in the most effective way and ensures a more thorough comparison of edits.
    \item[\textbf{D5.}] \textbf{Support verification of edit suggestions} As AI-powered tools automate suggestions, prior work has explored verifying AI's recommendations~\cite{huh2023genassist, ferdowsi2024validating}. In video creation, automated edit suggestions can often lead to additional errors (\textit{e.g.,} trimming a clip leading to jump cut) and thus needs careful validation. Future work can explore automatically identifying and communicating such erroneous suggestions to users.
    \item[\textbf{D6.}] \textbf{Support management and customization of variations}
    Formative study participants who reviewed a small number of alternatives all mentioned that they wanted to further refine the edits as none of the versions fully aligned with their preferences. Scaling up the number of alternatives increases the chance of finding a preferred version but it also makes it more difficult to manage and select from a larger pool of variations.
    

    
\end{itemize}

\section{VideoDiff}
We present \sysname{} (Figure~\ref{fig:teaser}), a human-AI video co-creation tool designed to support efficient video editing with alternatives.
With \sysname{}, users can generate and review diverse AI recommendations for 3 different video editing tasks: making rough cuts, inserting B-roll images, and adding text effects (D3).
VideoDiff supports easy comparison between alternatives by aligning videos (D1) and highlighting differences using timeline and transcript views (D2, D4)
. Users can organize and customize edits by sorting, refining, and regenerating AI suggestions (D6). 

\begin{figure*}
  \centering
  \includegraphics[width=\textwidth]{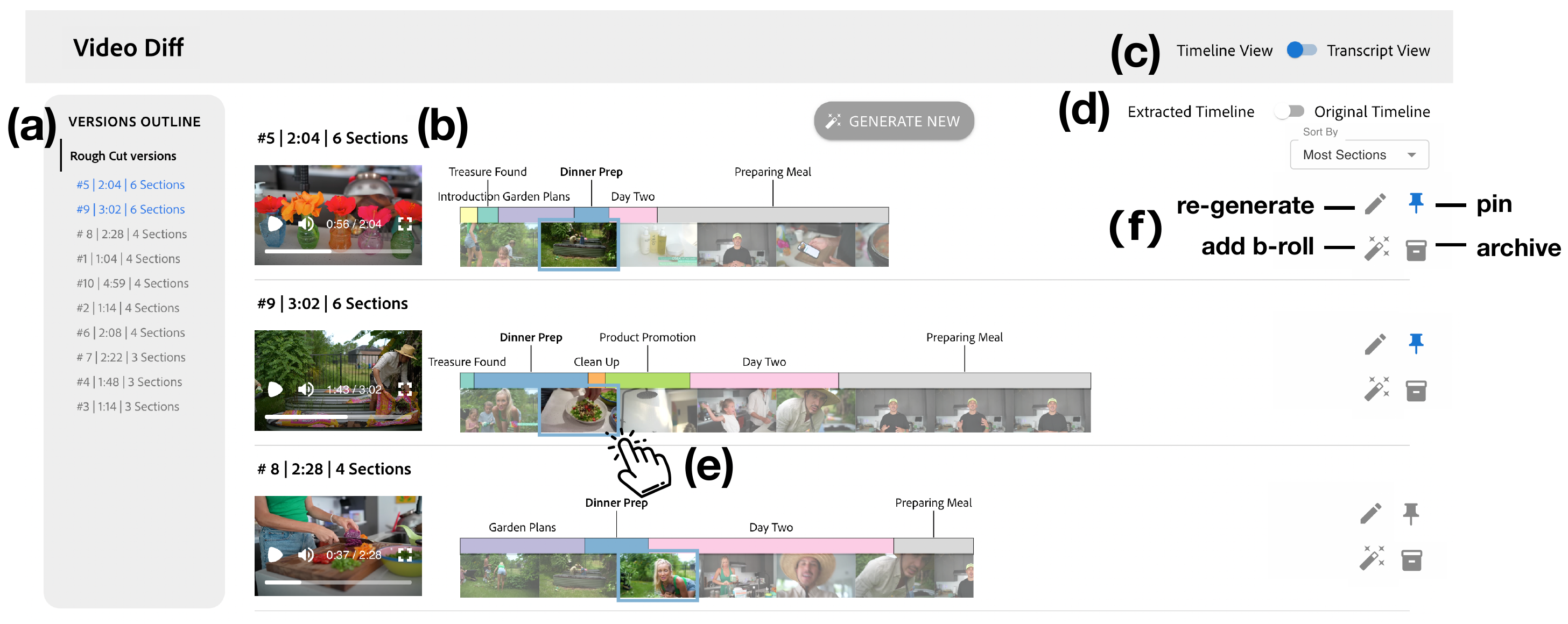}
  \caption{Overview of VideoDiff: Users can view an outline of the variations in the current editing stage (a). In this figure, we see 10 rough cut variations. The user can play videos of these different versions (b) and compare them in the transcript or timeline view (c).
  Users can toggle between the edited and source timelines (d) to align videos to the source or edited context or click on each section to navigate directly to that part of the video (e).
  Users can also sort variations by duration and the number of sections included, as well as pin, archive, or edit variations according to their preferences (f).}\label{fig:sys_overview}
\end{figure*}

\subsection{Scope}\label{sec:scope}
\revised{
While VideoDiff provides visualizations that allow users to quickly identify and skim differences between video versions (D2), it is designed to go beyond the functionality of a \textit{diff tool}, which typically focuses solely on highlighting changes or differences for review~\cite{tharatipyakul2018towards, baker2024interaction}. Instead, we designed VideoDiff as a human-AI co-creation tool that facilitates iterative \textit{generate-compare-refine} interactions, revealing how comparing alternatives can influence and enhance users’ creative processes.}

To help set context for \sysname{}, we define the different stages of video editing.
Rough cut creation is selecting good moments (or clips) from source footage to create a compelling story. Inserting B-roll images or video helps to make the video more interesting and dynamic. To insert good B-roll, creators should find effective images or videos that illustrate what is being said and place them appropriately. Inserting text effects helps emphasize parts of the narration through animation and stylized text. The AI algorithms for each of these editing tasks are in themselves hard technical problems, and we do not focus on them in this work. Our focus is on supporting users to work with the generated variations as part of the editing process. However, because we must support some editing in order to test our ideas, we have implemented basic versions of editing algorithms that leverage LLMs to process the video transcript and recommend rough cuts, B-roll, and text effects. A holistic solution with multimodal analysis of video, audio, and narration can offer better editing suggestions in the future.

Most video editing softwares offer more than the three stages that VideoDiff supports, such as applying color correction or cleaning up the audio. Our goal is not to create a fully functional video editor, but rather to explore how video editing might change as AI technologies enter more and more of the editing stages. \camready{As AI easily generates multiple editing recommendations, the video editing task shifts to involve more curation beyond just editing, and we must consider how to best support this transition.}

\subsection{Interface}
VideoDiff is a web-based video editing tool (Figure~\ref{fig:sys_overview}) where users can generate, review, and customize video alternatives. 
When the user uploads a video, \sysname{} first generates 10 rough cut recommendations. In the \textit{versions outline} on the left, users can see the list of all variations in the current editing stage (Figure~\ref{fig:sys_overview}.a). Users can play the video of each version (Figure~\ref{fig:sys_overview}.b) or skim the differences using a timeline or transcript view (Figure~\ref{fig:sys_overview}.c).  Users can toggle between showing only the edited or all of the source content for additional context (Figure~\ref{fig:sys_overview}.d).  They can also click on any section to navigate directly to that part (Figure~\ref{fig:sys_overview}.e).
%
Users can sort, re-order, pin and archive the variations to organize them (Figure~\ref{fig:sys_overview}.f). Or they can refine or recombine existing variations or regenerate a new variation using text prompts. 

\begin{figure}[t]
  \centering
  \includegraphics[width=\columnwidth]{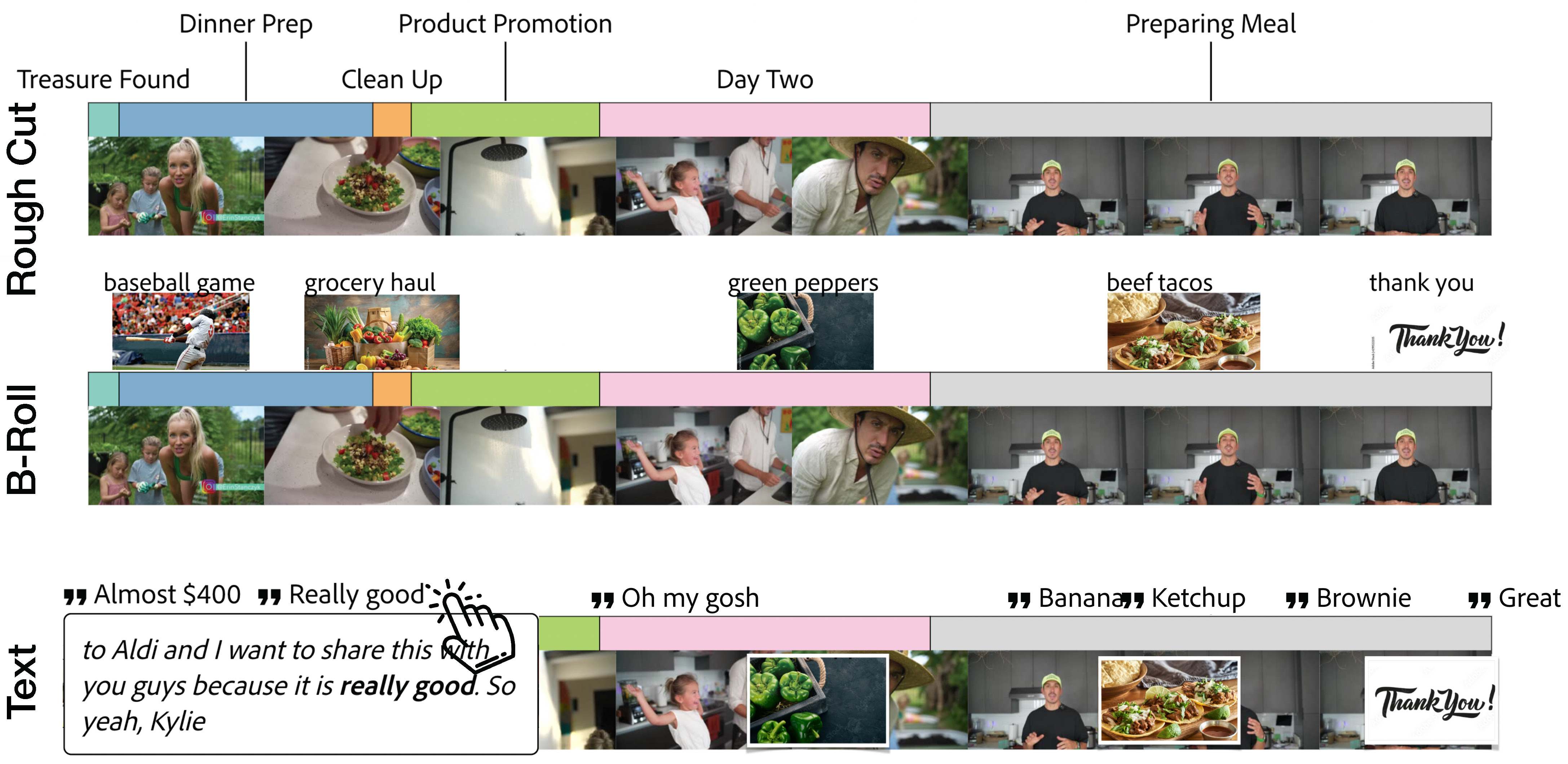}
  \caption{At each editing stage, VideoDiff provides glanceable timelines for users to easily compare different variations. Users can click on any section, B-roll image, or text effect to jump to that part of the video and preview the effects.}\label{fig:timelines}
\end{figure}

\begin{figure*}
  \centering
  \includegraphics[width=\textwidth]{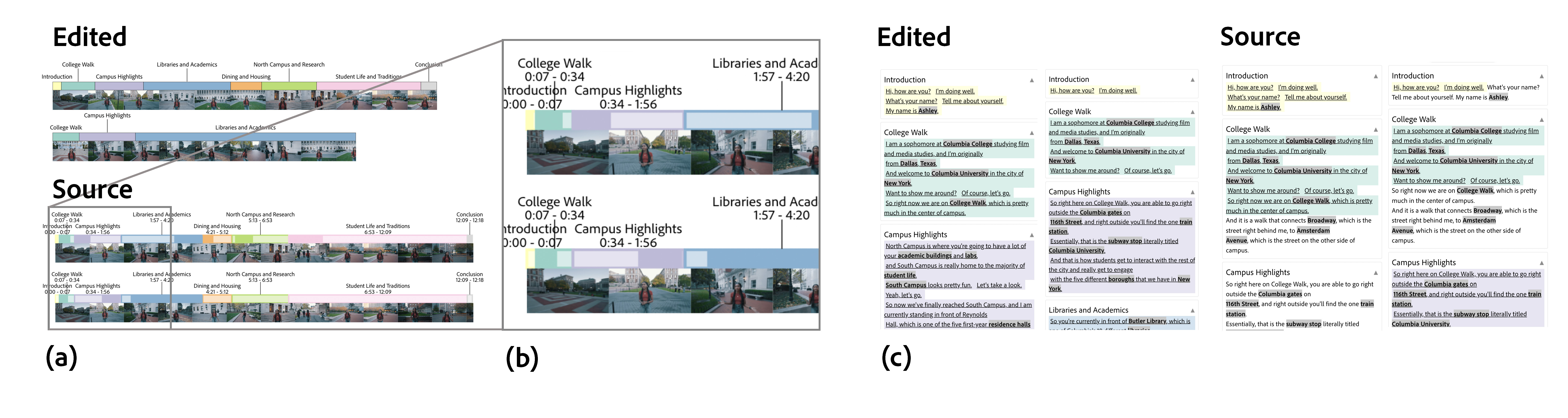}
  \caption{Users can switch between the edited and original source timeline in the transcript (a) and timeline (b) views. This helps users see the edits in the context of the source content and compare which sections are included at a glance. The source view (c) shows users the location of the edited content in the context of the source view.}\label{fig:extracted_original}
\end{figure*}

\ipstart{Comparing Variations with Timeline View}
VideoDiff provides different visualizations of timelines at each video editing stage (Figure\ref{fig:timelines}) so that users can quickly review and compare multiple alternatives (D3). 
In the rough cut stage, VideoDiff uses ~\textit{sections} to visualize the timing and coverage of content in each variation.
Drawing upon prior work that has shown that grouping footage into thematically coherent chunks can help creators make video editing decisions~\cite{leake2020generating, huh2023avscript}, we explore how chunking into sections can aid comparison of edited videos. 
Instead of segmenting each rough cut variation to identify sections, VideoDiff extracts sections from the source footage and applies them consistently across all variations. This allows users to easily align the variations and see how different versions include or exclude certain sections and cover varying parts or lengths of each (D1-D2). 
Users can toggle between edited and source timeline views (Figure~\ref{fig:extracted_original}.a) to align the videos based on the edited or source video timeline. The edited timeline allows users to quickly skim the video's overall duration, along with the placement and proportion of each section. By hovering over a section, users can view relevant video thumbnails and click to play the video from that point.
With the source timeline, users can easily understand which parts of the source video are extracted. In the formative study, video creators often compared edited videos to the source videos to verify any missing key information. Drawing from this observation and aligning with D1, we use the source video as an anchor for aligning rough cut variations for comparison. 
For example in Figure~\ref{fig:extracted_original}.b, while the edited timeline shows that both versions have a similar duration for ``Campus Highlights'' section, the source timeline reveals that each version covers different parts of the source footage on ``Campus Highlights''. By hovering over the lighter-colored boxes indicating excluded parts, users can see which video thumbnails and topic keywords are not covered.

Once the user chooses a rough-cut version, they can generate 10 videos with different B-roll recommendations. In the B-roll stage, VideoDiff's timeline view shows the B-roll thumbnails on top of the rough cut timeline bar, along with the keyword used to search for the B-rolls (Figure~\ref{fig:timelines}). Users can hover over the B-roll thumbnails to see which video scene each B-roll covers and click thumbnails to play the video and preview how the B-roll is inserted into the footage. After selecting a B-roll version, users can generate 10 new videos with different text effect suggestions. Instead of showing video thumbnails with text effects, we display the keywords where the text effects are applied, as text effects on thumbnails are too small to skim.
Users can hover over a keyword to view the narration sentence for context, and click on the keyword to play the video and preview how the text effects are integrated into the footage.

\begin{figure}[t]
  \centering
  \includegraphics[width=\columnwidth]{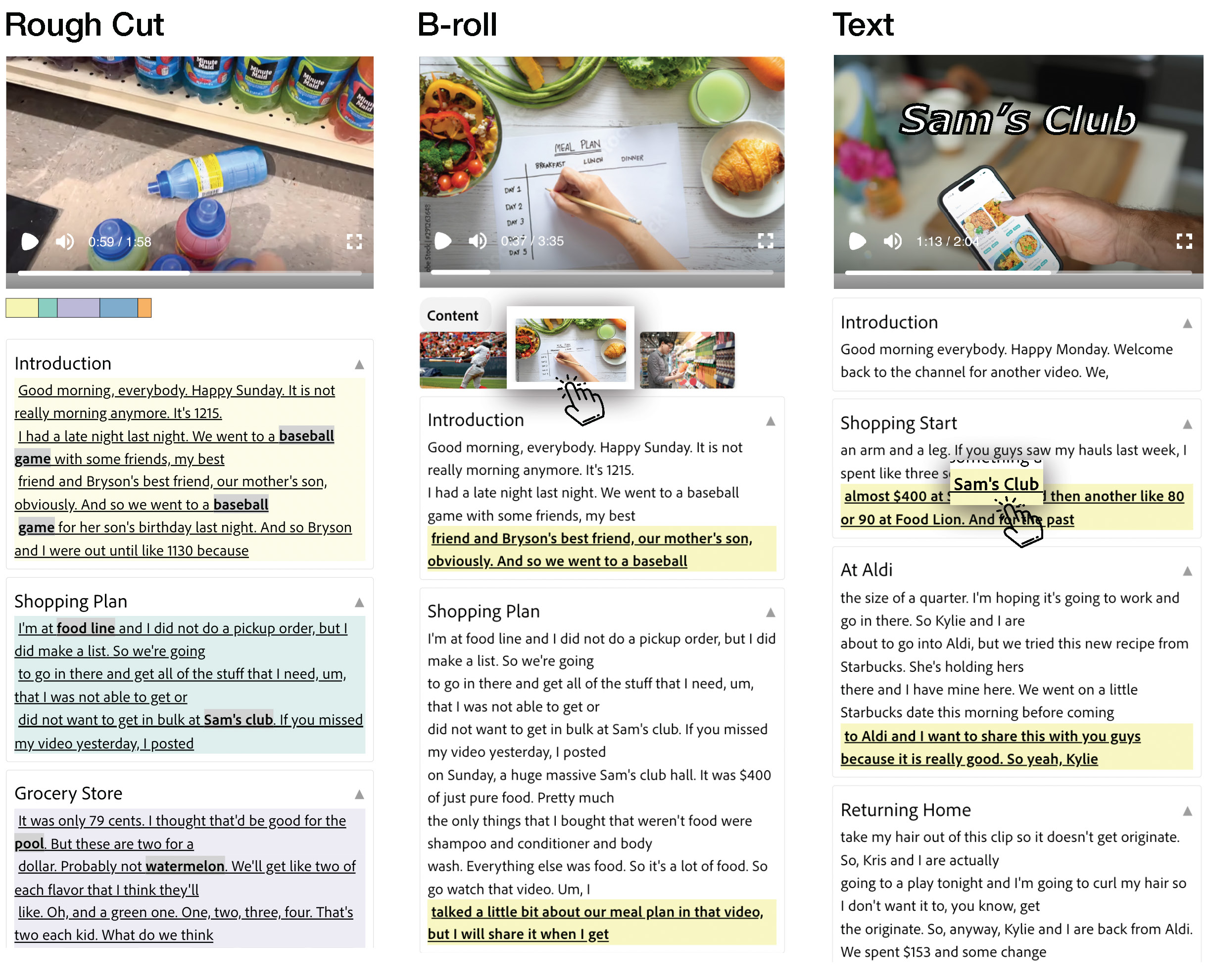}
  \caption{At each editing stage, VideoDiff provides glanceable transcripts so that users can easily review and compare different variations. In rough cut transcripts, visually concrete keywords~\cite{leake2020generating} are emphasized in bold, allowing users to easily skim through the content of each variation. Users can click on section headings, B-roll images, or text effects to jump to that part of the video and preview the effects.}\label{fig:transcripts}
\end{figure}

\ipstart{Comparing Variations with Transcript View}
VideoDiff's transcript view allows users to quickly skim the transcripts to understand the differences of variations in each editing stage (Figure~\ref{fig:transcripts}). 
The transcript is divided into sections and users can click on the section headings to play the video from the starting point of each section.
\camready{During the rough cut phase, VideoDiff aids users in maintaining context while toggling between transcript and timeline views by showing a mini timeline within the transcript view. It also uses consistent color coding to denote sections across both views.}
In the transcript, visually concrete keywords~\cite{leake2017computational} are emphasized in bold, allowing users to skim through the content of each variation~\footnote{We used GPT-o to identify visually concrete keywords with few-shot examples.}.
Similar to the timeline view, users can also switch to the source transcript view (Figure~\ref{fig:extracted_original}.c) and see the complete transcript and identify which parts of the source text are excluded in each section. We synchronize the scrolling across multiple transcripts, allowing users to easily compare variations side-by-side (D1). 

For quick skimming of B-roll and text effects options, VideoDiff highlights sentences in the transcript where these effects are applied (D2, Figure~\ref{fig:transcripts}). Users can click on the B-roll thumbnails or text effects to preview the video from the moment the effect appears.


\ipstart{Organizing Variations}
To help users to easily explore and narrow down the search space of alternatives, VideoDiff supports sorting of variations in each stage. In the rough cut stage, users can sort based on the edited video duration and the number of sections, and in the B-roll and text effects stages, users can sort based on the number of effects and the media type of B-rolls (image, illustration, video) or style of text effects (title, subtitle, lower thirds). 
Additionally, users can manually pin preferred versions to the top or archive unwanted versions to the bottom, which is also reflected in the outline view.

\begin{figure}
  \centering
  \includegraphics[width=\columnwidth]{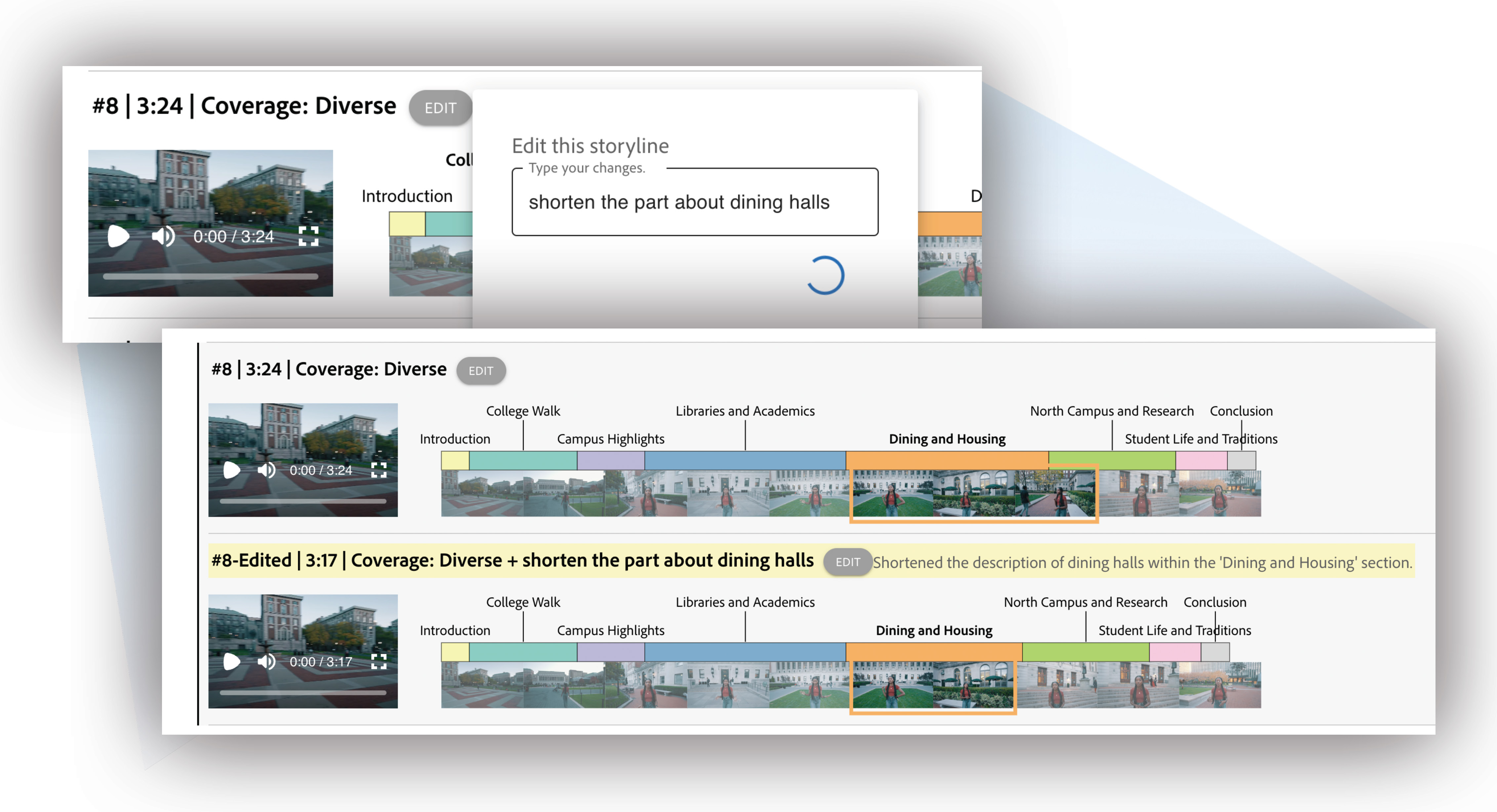}
  \caption{Using VideoDiff, users can edit a variation, recombine multiple variations, and generate a new variation using text prompts. For each new generation, VideoDiff summarizes the changes so that users can easily verify the result.}\label{fig:refine}
\end{figure}
    
\ipstart{Customizing Variations}
When users are not satisfied with the initial recommendations of VideoDiff, they can further edit existing versions or generate new alternatives (D6, Figure~\ref{fig:refine}).
Users can provide a text prompt to guide the generation of new alternatives (\textit{e.g., Show many text effects when I talk about grocery items.}), or click \textit{``Surprise me''} to have VideoDiff suggest a new alternative that is different from existing variations. 
Users can also recombine existing versions by specifying the version IDs and how to merge them in the text prompt (\textit{e.g., Use first two B-roll images from \#3 and last B-roll image from \#7.}). By default, generated new results are pinned to the top for easy discovery.
Using VideoDiff, users can also edit existing versions with a prompt (\textit{e.g., Shorten the part when I'm talking about the meal plans.}). The new generation from the edit prompt is displayed right below the original version for quick comparison and VideoDiff also describes specific changes made for quick verification of edits (D2). For example, in Figure~\ref{fig:refine} the user tells VideoDiff to ``shorten the part about dining halls'' and the system responds:
``Shortened the description of dining halls within the Dining and Housing section''. 

\subsection{Implementation \& Prompt Engineering}
We implemented VideoDiff using React.js and d3.js. For embedding a video player, we used Remotion~\cite{remotion} to render the edited video and overlay B-rolls and text effects. 
When users upload a video, VideoDiff uses OpenAI's Whisper API to transcribe the video. VideoDiff is powered by OpenAI's GPT-4o and uses prompt engineering for 1) segmenting video into sections and identifying visually concrete keywords, 2) generating multiple alternatives of edit recommendations for rough cuts, B-rolls, and text effects, and 3) parsing and executing users' new generation prompts and summarizing changes. 
To ensure VideoDiff provides diverse edit recommendations in each stage, we use ~\textit{augmentation prompts} (\S\ref{apndx:augmentation_pipeline}) to control the generation of suggestion (\textit{e.g.,} by specifying duration and section coverage for each rough cut recommendation). 

\section{User Evaluation}~\label{sec:user_eval}
To understand how well \sysname{} assists users in comparing multiple AI-edited videos, we conducted a within-subjects study with 12 video editors comparing \sysname{} to a baseline. 
Our study aims to investigate the following research questions:

\begin{itemize}
    \item[\textbf{RQ1.}] \textbf{How well does VideoDiff support video comparison?}
    \item[] Compared to the baseline, VideoDiff significantly decreases the time in video comparison (H1), improves comprehension and accuracy in video comparison (H2), lowers cognitive load (H3), and is more useful for video comparison (H4).
    
    \item[\textbf{RQ2.}] \textbf{How well does VideoDiff support video authoring?}
    \item[] Compared to the baseline, VideoDiff significantly increases satisfaction in the final video (H5), provides better creativity support (H6), and is more useful for video authoring (H7).
    
    \item[\textbf{RQ3.}] \textbf{How does VideoDiff impact video editing workflows?} 

\end{itemize}

\subsection{Method}
\ipstart{Participants}
We recruited 12 participants with diverse video editing experiences using mailing lists (P9-P20, Table~\ref{tab:participants}). 6 of them (P9-P14) described themselves as proficient, having 8.33 years of video editing experience (SD=2.94). The other 6 participants (P15-P20) identified as beginners with 3.83 years (SD=1.17) of experience. We compensated \$75 to professional video editors and \$30 to amateur editors for the 1.5-hour remote study conducted via Zoom.

\begin{figure}[t]
  \centering
  \includegraphics[width=\columnwidth]{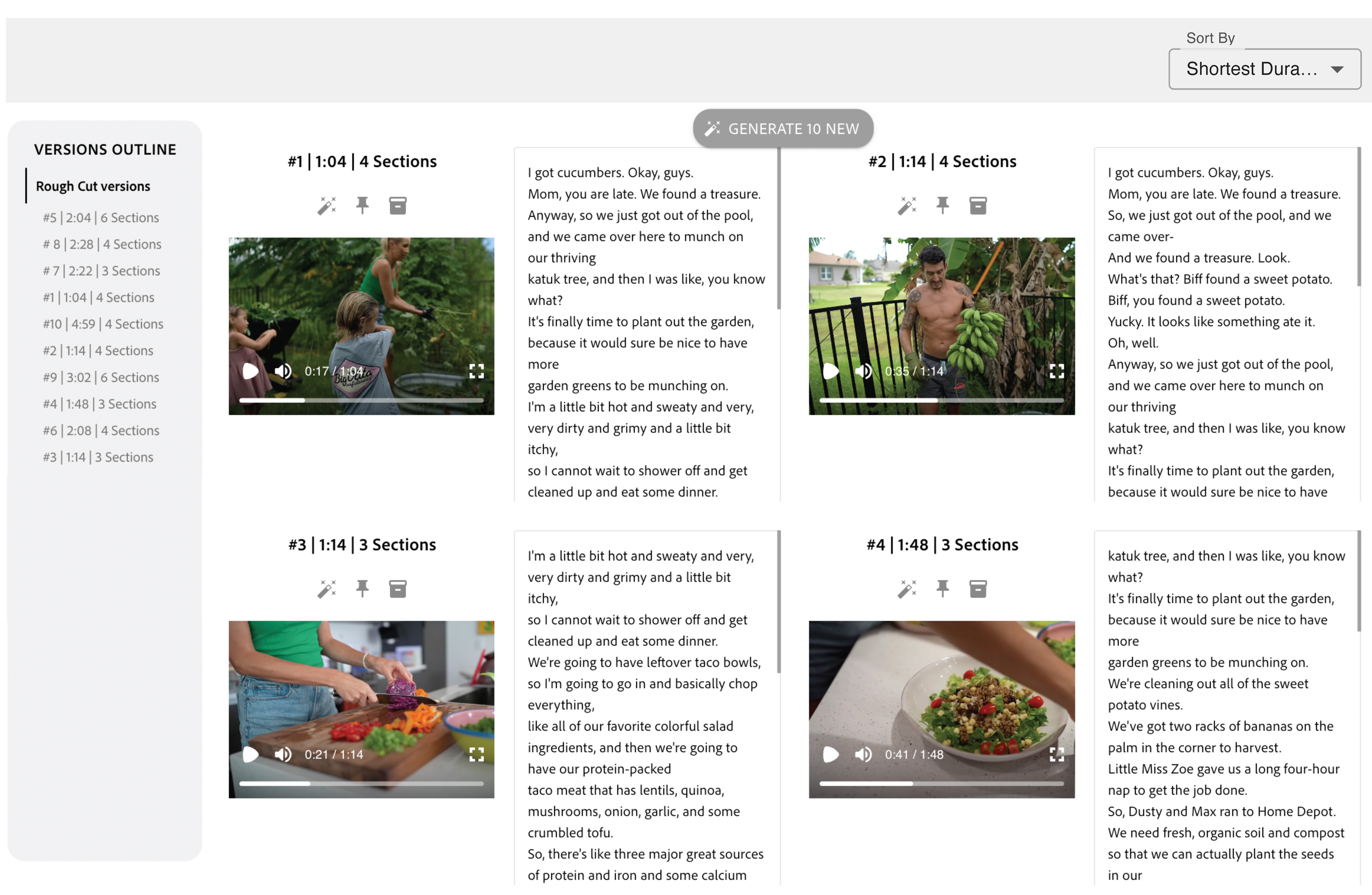}
  \caption{The baseline interface shares a similar UI design and features with existing AI video editing tools that support generating multiple videos (\textit{e.g.,} OpusClip, CapCut).}\label{fig:baseline}
\end{figure}

\ipstart{Baseline}
The baseline (Figure~\ref{fig:baseline}) shares a similar UI design and features with existing AI video editing tools that support generating multiple videos (\textit{e.g.,} OpusClip, CapCut). When the user uploads source footage, the baseline system generates 10 different videos with different rough cuts and B-rolls applied. Users can sort the clips based on the duration and the number of effects. 

\ipstart{Materials}
We selected 3 videos (V0-V2, Table~\ref{tab:video_materials}) from YouTube that feature primarily raw footage with few edits, real-world camera footage rather than screen recordings, and have narration. V0 was used in the tutorial session for both VideoDiff and baseline conditions. 
For the main study, we selected two videos (V1-V2) from the same YouTube channel which are similar in terms of length, amount of narration, and scene changes. We asked participants to watch and familiarize with the original videos prior to the study.

\begin{table}[h!]
\sffamily\def\arraystretch{0.9}\setlength{\tabcolsep}{0.4em}
  \centering
  \begin{tabular}{ccccc}
    \hline
    \textbf{Video ID} & \textbf{Video Type} & \textbf{Duration} & \textbf{Task} & \textbf{URL} \\ \hline
    V0 & Campus Tour & 12:28 & Tutorial & \cite{v0} \\ \hline
    V1 & Grocery Haul & 12:41 & Comparison & \cite{v1} \\ \hline
    V2 & Grocery Haul & 13:10 & Comparison & \cite{v2} \\ \hline
  \end{tabular}
  \caption{Videos used in the evaluation study.}
  \label{tab:video_materials}
  \vspace{-3pt}
\end{table}

\begin{table*}[h!]
\sffamily\def\arraystretch{0.8}\setlength{\tabcolsep}{0.4em}
  \centering
  \begin{tabular}{lll}    
    \toprule 
     Edit Stage & Category & Example Questions \\
    \midrule
    \multirow{5}{*}{Rough-cut} 
     & \multirow{3}{*}{Single-select} & (audio) Q1. \textit{Which video talks about most grocery items?} \\
     & & (audio) Q2. \textit{Which video mentions the calories of strawberry lemonade?} \\
     &  & (visual) Q3. \textit{Which video features the highest proportion of talking heads?} \\
     \cmidrule{2-3}
     & \multirow{2}{*}{Multi-select} & (audio) Q4. \textit{Which videos mention her salad recipe?} \\
     &  & (visual) Q5. \textit{Which videos show the scissors?} \\
    \midrule
    \multirow{5}{*}{\makecell{B-roll \&\\Text effects}} 
     & \multirow{3}{*}{Single-select} & (visual) Q6. \textit{Which video has the most B-rolls showing food?} \\
     &  & (visual) Q7. \textit{Which video shows the B-roll image when talking about the protein bar?} \\
     & & (both) Q8. \textit{Which video shows a B-roll image when talking about Z-bar?} \\
     \cmidrule{2-3}
     & \multirow{2}{*}{Multi-select} & (visual) Q9.\textit{Which videos show B-rolls with hamburgers?} \\
     &  &  (both) Q10.\textit{Which B-rolls are shown when talking about the restaurant?} \\
    \bottomrule
  \end{tabular}\caption{Questions used in the user study's comparison task included both single-select (choosing one answer that meets the criteria) and multi-select (selecting all answers that meet the criteria) formats. These questions required participants to check either the narration, visuals, or both, with the specific questions varying depending on the assigned video.}\label{tab:comparison_questions}
\end{table*}

\ipstart{Procedure}
We started the study with demographic and background questions about the participants' current video editing practices. Next, we gave a 10-minute tutorial on both \sysname{} and the baseline interface using V0. The participants then completed two tasks: the \textit{comparison task} (\textbf{RQ.1}) and the \textit{authoring task} (\textbf{RQ.2-RQ.3}). \revised{The comparison task studies how users use \sysname{} when the comparison is an end goal in itself and the authoring task explores when the comparison is the means to accomplish the higher-level goal of creating the final video.} 
In the comparison task, participants reviewed 10 different videos using \sysname{} or baseline to answer comparison questions (Table~\ref{tab:comparison_questions}). \revised{We derive these questions from the formative study to consider what aspects video editors compare to select the video they preferred: the content of the videos (\textit{e.g.,} which key visual scenes are included or which key concepts are mentioned) and the editing techniques applied (\textit{e.g.,} the style, placement, and frequency of visual effects). 
We aimed to cover diverse types of tasks, including both single-select and multi-select formats, as well as questions that required participants to assess narration, visuals, or a combination of both.
Also, our questions cover both comparison modes identified by Gleicher et al.~\cite{gleicher2011visual}: ~\textit{explicit comparison} where targets are known and available (\textit{e.g.,} Q4. \textit{Which video mention her salad recipe?}) and ~\textit{implicit comparison} where targets are hidden, requiring more exploratory search (\textit{e.g.,} Q3. \textit{Which video features the highest proportion of talking heads?}).} 
In each interface condition, users were assigned to one of V1-V2.
For each question type, we measured the time taken, answer accuracy, and interaction logs to understand which system components participants utilized to perform each comparison. For accuracy, we assigned a rating of 1 if the answer was completely correct, 0.5 if the participant got half or more of the answers right for multi-select questions, and 0 if less than half or none were correct.
After each interface, we conducted a post-stimulus survey that included the following ratings: mental demand, performance, effort, frustration, and usefulness of the system in understanding differences between videos. All ratings were on a 7-point Likert scale.

In the authoring task, participants reviewed the video suggestions and further edited videos using VideoDiff or baseline. In each interface condition, users were assigned to one of V1-V2.
We guided participants with an editing scenario: \textit{``You have created this initial video about grocery haul. To edit this video you can review AI's 10 different edit suggestions to select the final video or regenerate, refine, or recombine edit suggestions.''} After each interface, we conducted a post-stimulus survey to measure users' level of satisfaction with the final video, creativity support indexes (exploration, engagement, effort/reward tradeoff, expressiveness)~\cite{cherry2014quantifying}, and usefulness of the system in video authoring.
In both the comparison and authoring tasks, the order of the interfaces and videos was counterbalanced and randomly assigned to participants.

\subsection{Comparison Task Results}~\label{sec:comparison_task_results}
\begin{figure*}
  \centering
  \includegraphics[width=\textwidth]{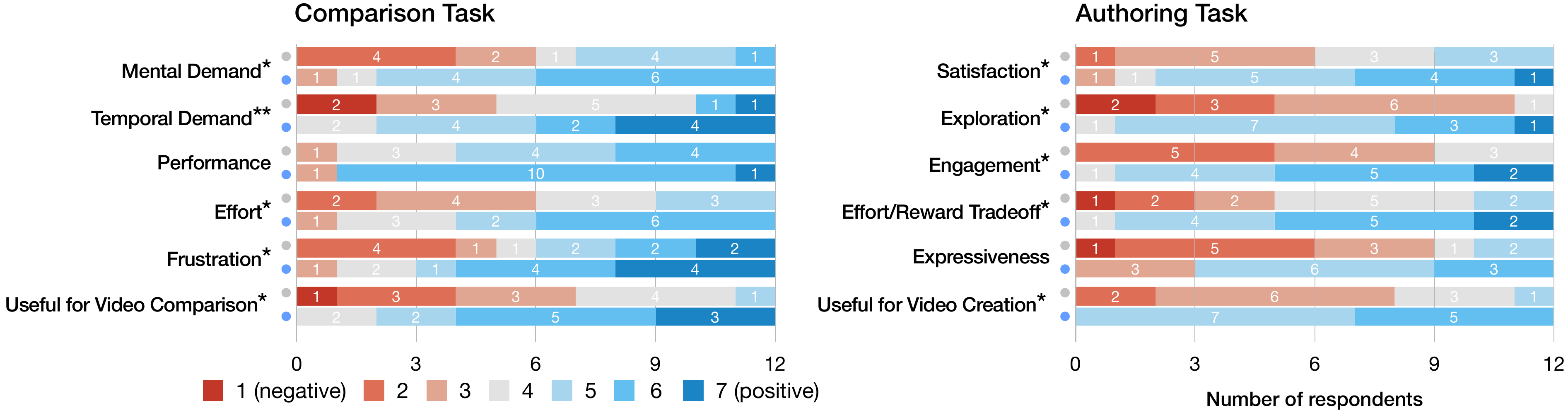}
  \caption{Distribution of the rating scores for the Baseline \raisebox{-1pt}{\includegraphics[scale=0.10]{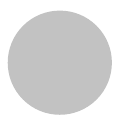}} and VideoDiff \raisebox{-1pt}{\includegraphics[scale=0.10]{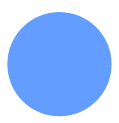}}
 (1 = negative, 7 = positive) in the two tasks. Note that a lower value indicates positive feedback and vice versa. 
  {The asterisks indicate the statistical significance as a result of Wilcoxon text} (~\textit{p} < 0.05 is marked with * and ~\textit{p} < 0.01 is marked with **). }\label{fig:survey_results}
\end{figure*}

Overall participants rated VideoDiff to be significantly more useful in understanding differences (Figure~\ref{fig:survey_results}, $\mu$=4.92, $\sigma$=1.16 vs. $\mu$=2.25, $\sigma$=1.06; $Z$=-2.77; $p$<0.05). In this section, we share 1) participants' strategies for reviewing and comparing videos to answer comparison task questions, 2) task results (RQ1), and 3) perceived workload (RQ1).

\ipstart{Comparison Strategies}
To narrow down the search space, 5 participants in the baseline condition (P9, P11, P14, P17, P19) and 6 participants in VideoDiff condition (P11, P14, P17-P20) first sorted variations based on the comparison question (\textit{e.g.,} sort by longest duration for Q1). With VideoDiff, 2 participants first identified which section is relevant to the question (\textit{e.g.,} ``Meal Plan'' section for Q9) then only checked within the variations that covered the section (P9, P18).
All participants using VideoDiff switched between timeline and transcript views depending on the question type. They used the edited view to find what each variation covers, and 2 participants additionally utilized the source view to align videos side by side (P10, P19). 3 participants were initially unsure which view would be more helpful for a specific comparison question and tried switching views multiple times (P12, P14-P15).

To answer the questions that require visual comparison between videos, all 12 participants in the baseline condition frequently interacted with the video player of each variation directly. Participants scrubbed the video player timeline to skim through the visual content of the video or click-jumped to navigate to different parts of the video.
After finding the visual of interest in one video, 4 participants tried to navigate to the similar timestamp in other videos but realized they needed to re-search as all videos extract different parts of the video.
As searching in each variation took time, P12 took notes to remember the details in each variation. She noted ~\textit{``10 is a lot to compare at once. I need to take notes.''}
Because participants could not preview thumbnails of multiple videos in parallel, 2 participants tried playing multiple videos synchronously but found it challenging to compare them as they were not aligned (P13-P14). As it was time-consuming to review the visuals of each variation, 4 participants used a workaround by searching for relevant keywords in the transcript (P14, P17-P19). P19 searched ~\textit{``I got ''} to check what grocery items are mentioned in the narration. 

Participants using VideoDiff mainly checked the timeline view to identify the shots included in each variation for rough cut comparison. If a timeline did not display a thumbnail of the visual they were searching for, 3 participants further ensured that the missing thumbnail was not simply due to the periodic nature of frame extraction while the visual is actually present in the video. P9 and P15 located sections that have thumbnails with similar background and shot type, navigated to the part and scrubbed the video player to skim the visuals. Similarly, P19 reviewed the ~\textit{source timeline view} to check whether the frame of interest was present outside of selected sections. 
When looking for specific B-roll images, P18 mentioned the benefit of visual keyword search using the B-roll keywords in the timeline view. To check the video context when a B-roll or text effect was applied, participants either reviewed the thumbnails below the effects or clicked on the effects to preview them.

For questions that involve checking the audio (\textit{i.e.} narration) of the video, participants in both conditions used keyword search (\smallverb{ctrl-F}) in the transcript. 
In the baseline condition, 5 participants read the transcript of each variation sequentially which was time-consuming (P9, P11, P14, P18-P19). P19, a low-vision participant noted ~\textit{``In this version [Baseline], it is mostly reading than watching so very tiring for my eyes!''}
With VideoDiff, 2 participants first searched where the relevant part appears in the transcript in one of the variations, then scrolled horizontally to quickly compare whether other variations cover the part using ~\textit{source view} (P9, P14).

\begin{figure}
  \centering
  \includegraphics[width=\columnwidth]{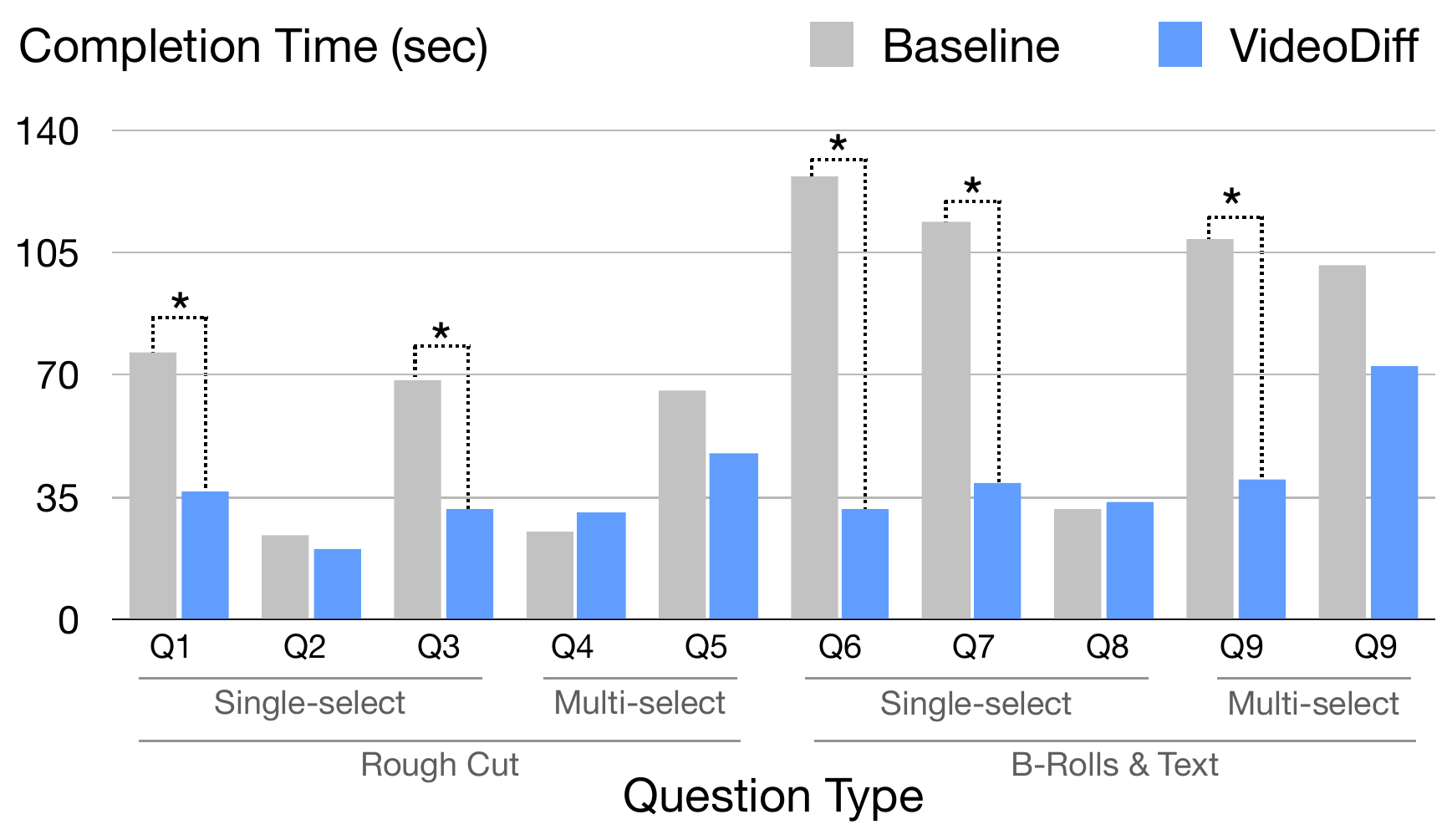}
  \caption{Average task completion time to answer comparison questions in Table~\ref{tab:comparison_questions}.(~\textit{p} < 0.05 is marked with *)}\label{fig:completion_time}
\end{figure}

\ipstart{Time and Accuracy for Comparison Task}
We report the results of the comparison task in Figure~\ref{fig:completion_time} (task completion time) and Table~\ref{tab:comparison_accuarcy} (task accuracy). On average, participants spent 38 seconds (SD=14) for each comparison question with \sysname{}, roughly half the time spent with the baseline (74 seconds, SD=38). In 5 questions that required checking multiple parts of the video (\textit{e.g., Which video has the most B-rolls showing food?}), participants using VideoDiff were significantly faster than in Baseline. 
In the baseline, 8 participants failed to find the correct answer for one or more questions within the 3-minute limit (P9, P12-13, P15-P16, P18-P20). 5 participants provided a rough guess answer instead of carefully comparing variations after watching the first few videos (P9, P12, P16-P18). P16 notes ~\textit{``In only three minutes? I'll just have to guess as I cannot watch all these videos.''}

Baseline participants often provided incorrect or incomplete answers when the question involved visual comparison. 3 participants who scrubbed the timeline to search for an object missed it when it only appeared for a short amount of time (P11, P17-P18). While 2 participants searched keywords related to the visual in the transcript to speed up the search, some videos only mentioned the object and did not show it (P12, P17).
While VideoDiff's filmstrip in the timeline enabled participants to more quickly compare the visuals of multiple variations, 5 participants missed the visual search question (Q5) when the object asked only appeared briefly in the video and was not captured in the filmstrip.

\begin{table*}[h!]
\small 
\sffamily\def\arraystretch{1.2}\setlength{\tabcolsep}{0.3em}
  \centering
  \begin{tabular}{c|cccccccccc}
    \hline
    \textbf{} & \textbf{Q1} & \textbf{Q2} & \textbf{Q3} & \textbf{Q4} & \textbf{Q5} & \textbf{Q6} & \textbf{Q7} & \textbf{Q8} & \textbf{Q9} & \textbf{Q10} \\ \hline
    \textbf{Baseline} & 0.67 (0.49) & 1.00 (0.00) & 0.42 (0.51) & 0.96 (0.14) & 0.63 (0.31) & 0.50 (0.52) & 0.36 (0.50) & 0.92 (0.29) & 0.46 (0.40) & 0.67 (0.25) \\ 
    \textbf{System} & 0.92 (0.29) & 1.00 (0.00) & 1.00 (0.00) & 1.00 (0.00) & 0.75 (0.34) & 1.00 (0.00) & 0.90 (0.32) & 0.91 (0.30) & 0.96 (0.14) & 0.86 (0.23) \\ \hline
  \end{tabular}
  \caption{Task Accuracy for answering comparison questions in Table~\ref{tab:comparison_questions}. The value represents the average accuracy (SD).}
  \label{tab:comparison_accuarcy}
\end{table*}

\ipstart{Cognitive Load in Video Comparison}
Figure~\ref{fig:survey_results} shows the distribution of survey ratings. Our NASA-TLX~\cite{hart2006nasa} results indicate that 
VideoDiff required significantly less mental demand ($\mu$=4.33, $\sigma$=1.4 vs. $\mu$=2.75, $\sigma$=0.97; $Z$=-2.08; $p$<0.05), temporal demand ($\mu$=4.33, $\sigma$=1.72 vs. $\mu$=2.33, $\sigma$=1.15; $Z$=-2.36; $p$<0.01), effort ($\mu$=4.42, $\sigma$=1.08 vs. $\mu$=2.92, $\sigma$=1.08; $Z$=-2.62; $p$<0.05), and frustration ($\mu$=3.75, $\sigma$=2.01 vs. $\mu$=2.33, $\sigma$=1.37; $Z$=-1.78; $p$<0.05). We did not see any significant difference in performance ($\mu$=3.08, $\sigma$=1 vs. $\mu$=2.17, $\sigma$=0.94; $Z$=-1.66; $p$>0.05). 
Participants using baseline mentioned the challenge of finding relevant parts with a video using a video player and transcript and having to repeat the process for multiple variations. P15 noted ~\textit{``It is more tedious than demanding. All videos are not visually distinct so I have to do a lot of manual digging.''} 

After using VideoDiff, participants appreciated having separate views to compare videos across timelines and transcripts. P11 mentioned the benefit of having both in separate views ~\textit{``If I'm shown both views at once, it is too overwhelming and will not fit multiple versions in a page.''} P10, P13, and P19 highlighted that the same color coding of sections across timelines and transcripts helped them pick up information easily. P13 said \textit{``While this one [VideoDiff] has much more information with many different view options, they are not overwhelming and all very helpful and complementary.''} P10 expressed that she wants to use VideoDiff in the future to show her clients multiple versions of videos she has created. 

\subsection{Authoring Task Results}
Overall participants rated VideoDiff to be significantly more useful for creating videos (Figure~\ref{fig:survey_results}, $\mu$=3.25, $\sigma$=0.87 vs. $\mu$=5.42, $\sigma$=0.51; $Z$=2.97; $p$<0.05). In the followings, we share 1) participants' strategies for reviewing and customizing variations for creating a video (RQ3) 2) how well VideoDiff supports creativity (RQ2) and 3) future improvements suggested by participants.

\ipstart{Reviewing Variations and Converging}
In the baseline condition, participants mentioned the difficulty of understanding the differences and quickly identifying which one aligns best with their preference. Thus, 3 participants quickly selected the longest video as they were likely to cover most of the original video (P12-P13, P16). P16 mentioned ~\textit{``I cannot understand what each version is about, so I'll choose the longest one and import it into my own video editor to cut down further.''} Similarly, when reviewing the baseline's B-roll recommendations, P18 chose a version with the most B-rolls as it was easier to remove than add them manually. 
4 participants who first edited videos with the baseline mentioned feeling overwhelmed by starting with 10 alternatives and wanted fewer variations (P10-P12, P14). However, when they used VideoDiff to create videos in the following sessions, P10 and P14 changed their opinion and stated that they liked having 10 variations as a start. P14 said \textit{``Because I can quickly skim the differences with visuals and zoom out, the same amount of suggestions don't feel overwhelming here. ''}

When creating videos with VideoDiff, all participants started in the timeline view and narrowed down the search. To closely compare final candidates or verify edits, 10 participants switched to the transcript view for a side-by-side comparison. P16 said~\textit{``You cannot zoom out and view all text in the transcript at once. It's possible with the timeline view.''}
Similarly, P17 compared two views: \textit{``Timeline view is helpful for creating bones of the videos, deciding what to show for how long. Transcript helps with individual details.'' }
2 participants never used the transcript view and finished editing in the timeline view (P17, P19). P19 said ~\textit{``All video editing software that I use shows timeline, so I'm not used to editing with the transcript view.''} 

In the rough cut stage, participants sorted and shortlisted variations based on the coverage of sections they were interested in. 3 professional video editors highlighted the benefit of segmenting the video into sections (P9, P11, P14). P14 mentioned~\textit{``This is exactly what I do manually every time I start video editing, using colors to differentiate my shots in Premier Pro. This system [VideoDiff] will save me so much time when I'm working on 2-3 hour-long videos.''} 
P15 who narrowed down the rough cut versions to 2 candidates first added B-rolls to both versions to inform decision-making in the rough cut stage. She explained~\textit{``If one version has a part that I can add nice B-rolls, I'll go for that [Rough cut version].''}
When reviewing B-roll and text effect recommendations, participants checked whether the effects were evenly distributed throughout the video using timeline view (P12, P17), whether the B-roll aligns with the narration in the transcript view (P9, P11), and whether the B-rolls and text effects are visually aesthetic and match the style of the video by previewing in the video player (P13).

\ipstart{Generating and Editing Variations}
In the baseline condition, only 2 participants generated a new set of variations (P11, P13) while the other 10 participants selected the final video from the initial recommendation without further change. P19 said ~\textit{``I wouldn't generate new videos because that would require another comparison!} After generating new recommendations, P11 noted ~\textit{``It is difficult to know how these videos are different from the original variations.'' }

Participants who created videos using VideoDiff made 4.33 edits to existing variations (SD=2.42). 
2 participants combined existing versions (P13, P19) and
3 participants generated new variations (P11, P16-P17). During the rough cut stage, many participants made edits related to sections, such as adding or removing sections, adjusting the coverage of sections, or reordering them. Participants made more edits focused on visual changes in the timeline view (P20 edited \textit{``Show less talking heads, show more grocery items''}), while they made more edits related to narration in the transcript view (P17 edited \textit{``Do not mention the store name''}). While most participants iteratively made edits on the first rough cut variation they selected, P18 switched to another version after making 3 edits on a rough cut version. He described~\textit{``After making a few edits on this one, I realized that I didn't really like it. It's really useful to explore multiple paths at once, and be able to switch between them easily.''}

In the B-roll and text effects stage, participants made edits to add/remove an effect, move the effect to a different part of the video by specifying a timestamp or a transcript sentence, and change the B-roll images or text styles. Participants noted that VideoDiff’s description of specific changes makes it easy to verify new changes without having to play the videos before and after the changes.



\ipstart{Video Diff as a Creativity Support Tool}
We assessed VideoDiff and the baseline using creativity support index~\cite{cherry2014quantifying} (Figure~\ref{fig:survey_results}). 
VideoDiff supported efficient exploration and comparison between variations ($\mu$=2.5, $\sigma$=0.9 vs. $\mu$=5.33, $\sigma$=0.78; $Z$=3.07; $p$<0.05), which enabled participants to quickly narrow down the search and spend more time customizing them. This led to higher effort--reward tradeoff ($\mu$=3.42, $\sigma$=1.24 vs. $\mu$=5.67, $\sigma$=0.89; $Z$=2.89; $p$<0.05) and higher satisfaction in the final result in VideoDiff than in the baseline ($\mu$=3.67, $\sigma$=0.98 vs. $\mu$=5.25, $\sigma$=1.06; $Z$=2.63; $p$<0.05). After creating the final video with VideoDiff, P13 said ~\textit{``I can see that this version covers all sections in a balanced way and has many aesthetic B-rolls, so I'm confident that I chose the right one.''}

Participants were also more engaged in the creative process with VideoDiff than the baseline ($\mu$=2.83, $\sigma$=0.83 vs. $\mu$=5.67, $\sigma$=0.89; $Z$=3.05; $p$<0.05). P15 compared the experiences in both conditions: \textit{``In the first session [Baseline], I had to read the redundant transcript over and over, and having to read most of the time when I'm trying to create a video is not so fun. With this one [VideoDiff], I didn't need to do repetitive work and could experiment what AI can do, so really enjoyed using it!''}
For expressiveness, we did not see any significant difference between two conditions ($\mu$=2.83, $\sigma$=1.27 vs. $\mu$=4.75, $\sigma$=1.14; $Z$=2.46; $p$>0.05). 
All participants using the baseline interface felt disengaged from the creative process due to limited control over the generation and a lack of understanding of the variations. P19 stated~\textit{``There was no creation in this process, I was just selecting like a judge.''} Similarly, P11 said~\textit{``While being able to generate 10 recommendations is impressive, I'm not providing any input so I don't feel like I'm the one creating something.''}
In the VideoDiff condition, participants showed mixed perspectives towards VideoDiff as a creativity support tool.
P11 described~\textit{``Instead of spending most of the time reviewing, it [VideoDiff] helped me to quickly move on to the creative stage and spend more time and effort there.''} P14 elaborated~\textit{``This tool [VideoDiff] helped me with tedious choices so that I can spend more time on meaningful choices.''}
On the other hand, P10 mentioned that VideoDiff does not make her feel as expressive as using her existing tools.~\textit{``While this is helpful for making things faster, it feels like I'm supervising AI instead of me being creative. I might feel less attached to what I create with this tool.''}


\ipstart{Future Improvements for VideoDiff}
Participants provided suggestions on how to improve \sysname{} in the future.
First, P13 and P20 suggested that VideoDiff can provide alternatives whenever participants request edits to existing variations. Current VideoDiff provides only one edited variation based on the user's edit prompt. Thus, when the new variation does not align with the user's intent, they have to refine the edit prompt multiple times. P20 described~\textit{``I cannot always be specific because I forget to specify all the details in the prompt or I often do not know what I want. When I ask it to ``shorten grocery haul'', it can generate 3-4 versions with different parts dropped.''}
Second, 5 participants (P9-P11, P14, P16) wanted more direct control with VideoDiff. P11 described~\textit{``While I can change the duration or change B-rolls using prompt-based editing, some of these are easier with directly stretching the timeline or right clicking on B-rolls.''} P9 also suggested having a slider to control the number of B-rolls and text effects easily. P16, an amateur video editor mentioned~\textit{``Instead of exporting this and further refining using Premier Pro, I just want to finish everything here and directly publish without having to work twice.''}
Third, P14 mentioned that future VideoDiff can improve its generation pipeline to consider visuals into account when generating rough cuts and inserting B-rolls. P14 explained~\textit{``I want B-rolls to be placed where the jump cuts are or when there are no interesting visuals. Now it covers the original video's bananas with a B-roll bananas.''}
Finally, P19 a low vision participant provided insightful feedback to make VideoDiff more accessible to blind and low vision video creators. For low vision users who may find reading transcript challenging, VideoDiff can include more detailed subheadings summarizing the transcript's content, in addition to the current section titles, to reduce the amount of reading required. For screen reader users, providing more descriptive text for visuals (\textit{e.g.,} alt-text that describes differences between 2 similar B-roll images~\cite{huh2023genassist}) can support accessible visual comparison of videos.

    
\section{Exploratory Case Studies}
The controlled user study (Section~\ref{sec:user_eval}) showed that video creators found it easier to comprehend differences and create videos using VideoDiff compared to the baseline. To learn how creators would use VideoDiff to edit their ~\textit{own footage}, we conducted an exploratory study with 3 video creators (P9, P21-P22) who brought their footage.

\begin{figure}[t]
  \centering
  \includegraphics[width=\columnwidth]{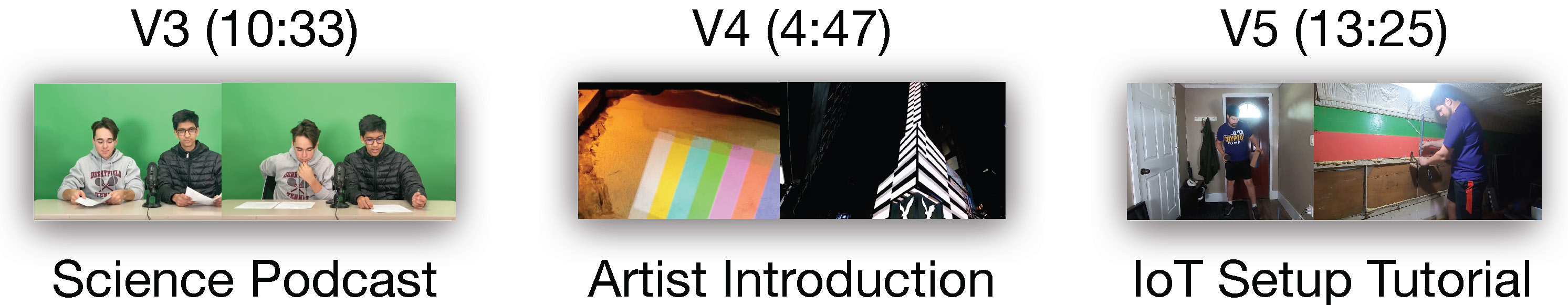}
  \caption{In exploratory case studies, creators edited their own footage (V3-V5) using VideoDiff.}\label{fig:exploratory_videos}
\end{figure}

\subsection{Method}
We recruited 3 video editors (2 professionals, 1 amateur) using mailing lists and social media (P9, P21-P22). P9 also participated in the controlled user evaluation study. Before the study, we collected footage from each participant (Figure~\ref{fig:exploratory_videos}). P9 and P22 who wanted to use multiple clips in the study concatenated them into a single source video.
During a 1hr remote study session, we asked participants background questions, provided a tutorial of ~\sysname{}, invited participants to edit their own footage with ~\sysname{}, and asked participants semi-structured interview questions about their experience. We compensated participants \$75 via Amazon Gift Card for preparing the footage and participating in the study.

\subsection{Three Vignettes: VideoDiff in Context}
\ipstart{V3: Science Video Podcast}
P21 is a college student who creates videos for a personal YouTube channel and school projects. For his physics class, he created a video podcast that parodies golf announcers to explain Newton's law in a fun way. His goal for editing was to make an informational and entertaining video while describing all three of Newton's laws. With VideoDiff's initial rough cut suggestions, he first shortlisted versions with many sections which are likely to cover all three laws. P21 noted ~\textit{``It'd be nice if I could ask it to filter out versions that do not cover all three laws''}. While he chose a longer rough cut version to further edit, he also pinned a shorter version, which primarily consisted of jokes, to use as a teaser for his main video.
When reviewing VideoDiff's B-roll recommendations, he looked for a version with a more consistent style of B-rolls as he was making an educational video. He liked having abundant options for B-rolls and recombined suggestions to create a version that has many infographic illustrations. 
For text effects, he wanted to insert a chapter text for each law but noticed that most versions focused on formulas or jokes. He generated a new version by specifying ~\textit{``Show a lower-third text of what each chapter is about.''}
In the future, P21 wished that VideoDiff would gradually learn from his selection of alternatives and offer more personalized alternatives. ~\textit{``Like reinforcement learning, I want it to learn my preferences over time so I can spend less time customizing.''}



\ipstart{V4: Artist Intro Video}
P22 is a professional video creator who often edits music/art videos and documentaries for clients. He is making an artist intro video with the footage he got from a client and used VideoDiff to make edits. His goal for editing was to fit the video into 2 minutes while talking about the major art pieces. 
When reviewing rough cut variations, P22 first noticed that \#7 and \#8 are both close to 2 minutes, and compare them side-by-side in the transcript to choose a version that mentions more art pieces of the artist.
While he tried to edit the video with a prompt ~\textit{``Remove shots with flashing lights.''}, he noticed that they still remained in the video.
In the B-roll stage, P22 did not apply any B-roll and explained~\textit{``I want to show the artist's own art images rather than stock images.''} He suggested that in the future, VideoDiff can recommend zoom and pan animations for B-roll, as well as movements for text effects, and display them side by side for comparison.
He wanted to continue using VideoDiff as part of his workflow as the alternatives gave him new ideas. ~\textit{``My brain starts to lock once I start manually editing. Even when I try to tweak to something different, it gets fixated easily. It will be always useful to have alternatives.''}



\ipstart{V5: IoT Setup Tutorial}
P9 is a freelancer video creator who edits videos for YouTuber's vlogs and product reviews. One of the YouTubers asked him to edit their raw footage of a IoT setup tutorial to remove mistakes and make a polished video. 
P9 first chose one rough cut version and tried to make a section longer by editing with a prompt~\textit{``Add more footage for hotspot installation.''}. After noticing that VideoDiff generated a too long version, he switched to the transcript view to check what specific sentences he want to add and refined the edit prompt. 
After reviewing the footage, P9 noticed a lot of repetition in the sentences as the YouTuber corrects themselves. 
He edited the video to remove the repeated sentences and also prompted VideoDiff to cut when the YouTuber is out of the frame. However, he noticed that VideoDiff hallucinated that it had cut the parts out when they were still present.
He mentioned that while VideoDiff is great at making a rough version of a video, he still needed to rely on other tools to manually remove filler words and polish the video. P9 described~\textit{``Usually making a rough cut for a video this long takes about 2 hours, but today I did it in 5 minutes. Although I still have to polish it up, this is so useful!'' }




\section{Discussion}
In this section, we reflect on our findings from the design and evaluation of VideoDiff. We also discuss future opportunities to improve and extend VideoDiff. 

\ipstart{Creating Videos with Alternatives}
VideoDiff is a tool designed to support human-AI co-creation of video with alternatives. Our focus on alternatives is driven by prior work that highlights the value of divergent thinking in creative process~\cite{suh2024luminate, dow2010parallel, reza2023abscribe}. Video creators in our user study highlighted 3 benefits of having alternatives. First, it helped to consider alternative edit styles that they are not familiar with. Second, when users give the AI an ambiguous edit request, having multiple alternatives allows them to choose the one that best matches their intent, rather than relying on a single suggestion from the AI that might be incorrect. Third, with alternatives, creators can quickly prototype multiple versions of end-to-end videos adapted to different audiences, settings, and platforms. 
\revised{While editing videos with alternatives is not a one-size-fits-all solution, it represents existing practices in video editing and other generative AI tasks. In this work, we focus on minimizing the burden of working with alternatives in video creation and demonstrate how VideoDiff achieves it.
Still, some users may prefer incremental editing approaches, such as refining a single video~\cite{wang2024lave}, or workflows that prioritize error correction over creative decision-making~\cite{huh2023avscript}. In the future, we will consider diverse workflows by letting users select the number of suggestions (one vs many) and further explore how the scale of alternatives or the timing of providing alternatives can impact users' creation process.
}

\ipstart{\revised{Scope and Use Cases of VideoDiff}}
\revised{We reflect on the scope of VideoDiff based on user groups, creation stage, and control granularity based on the taxonomy of roles that creativity support tools play~\cite{chung2021intersection}. 
First, VideoDiff can support a diverse range of users. Novices or creators with accessibility needs (\textit{e.g.,} screen reader users) who find existing video editing software complicated to use can benefit from VideoDiff’s suggestions. Novices with limited experience with the authoring tool often do not know where to start~\cite{ashtari2020creating} and find it easier to start with examples rather than from scratch~\cite{kim2015motif}. 
For experienced editors, VideoDiff reduces the time spent on routine tasks such as planning and making rough cuts, allowing them to focus on fine-grained adjustments that improve the overall quality of their work.}

\revised{In terms of the creation stage, VideoDiff primarily supports idea generation and execution but not evaluation, requiring users to review and evaluate the quality. While automating execution, VideoDiff still emphasizes user control, allowing for post-editing and customization through natural-language prompts.
In terms of control, VideoDiff provides indirect control (\textit{i.e.,} text prompt) to prioritize a low learning curve over precise editing. This design choice contrasts with traditional video editing software, which offers precise but complex editing capabilities. Thus, it will be more useful in cases when users want to create videos under time constraints (\textit{e.g.,} social media content, event recaps) or budget limits (\textit{e.g.,} personal projects or prototyping for feedback before detailed editing). While users can export the EDL to further edit in other video editing tools, future VideoDiff can incorporate more direct control (\textit{e.g.,} timeline manipulation, dragging B-roll to adjust timing).}


\ipstart{Support for Flexible Comparison}
VideoDiff enables quick comparison of video alternatives by supporting users to align multiple videos and highlighting differences using timeline and transcript views. 
While creators using VideoDiff were successful with various comparison tasks, 2 participants were confused about which views to utilize for a few comparison tasks. 
To further minimize creators' mental load in search and comparison, a future iteration of VideoDiff can directly take creators' natural language comparison queries (\textit{e.g., Which video has fewer narration mistakes in the intro of the video?}). It can extract relevant parts of each video and determine the best views to communicate the differences and present them to creators. This can further reduce creators' effort to switch views and help scale up the number of variations for comparison.
Also, VideoDiff's pipeline can consider broader types of comparison beyond what is based on narration and visual content -- shot types (\textit{e.g., Which video features a higher proportion of a talking head?}), audio (\textit{e.g., Which videos use calm music?''}, errors (\textit{e.g., Which video has the least sentence cut-offs?}), or abstract comparison queries (\textit{e.g., Which video has a more hooking intro?}) 
\camready{Also, while D4 highlights comparison through multiple modalities, current VideoDiff focuses on visual and narration differences and does not consider sound or music differences. Prior work has explored audio tagging of videos~\cite{zhang2023peanut} and generating sound effects to augment video skimming~\cite{ning2024spica}. Similarly, future VideoDiff can consider audio-based comparison by surfacing differences in background noises, music inserted, or speech characteristics.}
As the pipeline evolves to identify and describe these various types of differences, finding effective ways to visualize them becomes another important challenge.

\ipstart{Implications of VideoDiff on Creativity}
\revised{
While AI-powered creativity support tools risk design fixation~\cite{wadinambiarachchi2024effects} and lack of control~\cite{suh2024luminate}, we designed VideoDiff to mitigate these concerns. First, VideoDiff supports divergent thinking. Participants in our study mentioned that it is easy to get fixated on one video editing style with existing tools, whereas VideoDiff inspired them to explore alternative ideas, which they found valuable to incorporate.
Second, VideoDiff supports customization. Instead of generating a single result for users to accept as-is, it offers multiple options, enabling users to select and refine suggestions. By not delegating all edit decisions to AI, VideoDiff ensures creators remain in control of the final result. Finally, VideoDiff promotes efficiency by automating tedious edits, allowing creators to spend more time in the creative stages.}

\revised{To further support a deeper sense of agency and ownership, future iterations of VideoDiff can involve users in the planning and ideation stage~\cite{chen2019neural} and take user inputs before generating suggestions. For instance, users could specify requirements such as duration or style, provide text-based instructions, or upload reference videos to guide the editing process. They could also upload a partially edited video—for example, one with the first half completed—and request the system to finalize the remaining edits.
To support more flexible user control, future VideoDiff can allow users to directly accept or reject individual suggestions as in prior works ~\cite{laban2024beyond, han2020textlets}. To improve the transparency of AI suggestions, we can provide explanations for the generated suggestions~\cite{wan2024felt} or offer access to intermediate steps of LLM chaining used in generating them~\cite{wu2022ai}. This can enable users to better understand, utilize, and experiment with the tool~\cite{yeh2024ghostwriter}. 
Furthermore, future work can consider a personalized recommendation pipeline within VideoDiff, where the tool adapts to a user's unique style and preferences over time. This would not only speed up the editing process but also give creators a more tailored experience.} 

As AI continues to accelerate the video creation process, there is also a risk that users may not thoroughly review or verify all the generated content. VideoDiff can help mitigate this by encouraging creators to do more sensemaking before making final selections. By offering clear visualizations of differences and allowing users to explore how each version fits their specific goals or audience, VideoDiff can assist creators in making more informed decisions.

\ipstart{\revised{Environmental and Societal Impact}}
\camready{
We discuss the long-term impact of AI-powered creativity tools such as VideoDiff on the environment and society.
Like other AI tools, the language model that powers VideoDiff requires computational resources for training and inference which involves carbon emissions~\cite{weidinger2022taxonomy}. 
Yet, our study demonstrates that VideoDiff is highly valuable to users by streamlining complex editing tasks and enabling creative exploration. Also, VideoDiff may reduce the time users spend on traditional editing workflows, which often rely on resource-intensive video editing software and involv repeatedly rendering, storing, and transferring multiple large video files via cloud services. Instead, VideoDiff represents edit variations as structured text, eliminating the need to store multiple edited versions as full video files. We expect the financial and environmental costs of the models to change over time, and future work can explore how well smaller models can support tasks of suggesting video edits.}

\revised{Generative AI tools like VideoDiff also raise concerns about job displacement in fields that rely heavily on manual video editing and production. However, VideoDiff does not automate the entire video editing process and retains key editing decisions with users who select and refine the suggestions. For novices, VideoDiff can lower the entry barrier to video editing and open doors for new roles. For experienced users, it reduces the time spent on tedious, repetitive tasks, allowing them to focus more on higher-order creative endeavors.
By fostering both accessibility and efficiency, VideoDiff can promote collaboration between humans and AI in a way that creates opportunities.}





\ipstart{Limitations}
VideoDiff demonstrates the usefulness of video creation with alternatives in 3 representative video editing tasks: making rough cuts, inserting B-rolls, and adding text effects. 
Our study participants expressed excitement to use VideoDiff for a wider range of edits (\textit{e.g.,} motion graphics, and transitions). Future versions of VideoDiff could incorporate more diverse editing types.

While VideoDiff currently takes a single source video as input, study participants mentioned that VideoDiff will be useful for generating and comparing alternative curations of multiple clips when provided with a huge library of footage. To support this, we need to explore ways for users to efficiently skim through the coverage and sequence of different clips, while also accounting for the possibility that a single clip may appear multiple times in the final edit.

VideoDiff is powered by an LLM (GPT-4o) to generate edit recommendations and to follow users' commands to generate new variations. Our pipeline is prone to transcription errors and LLM hallucinations as few study participants have pointed out. GPT-4o also has a limit for input tokens (128k) and cannot process long videos with extensive length of transcripts.
Also, the current pipeline does not consider visual input when generating edit recommendations and thus cannot correctly follow edit requests related to the visual content. In the future, large multimodal models that take visual input (\textit{e.g.,} GPT-4V) can be integrated to improve the performance of the pipeline. 

\section{Conclusion}
We presented \sysname{}, a human-AI video co-creation tool designed to support efficient video editing with alternatives. Through a functional prototype and quantitative and qualitative user feedback, we showed that users want to work with variations when they are editing videos and would like flexible tools that offer easy ways to minimize watching and maximize creative potential. \camready{VideoDiff inspired users to explore alternative ideas while retaining key edit decisions to users who can select and refine AI suggestions.
All study participants wanted to integrate it into their workflow to create more satisfying videos.} Generative AI tools offer new promising capabilities that will only improve over time. It is up to the HCI community to create new experiences that harness that power into better and better creative tools.
\begin{acks}
We thank Columbia College for giving permission to use the video \href{https://www.youtube.com/watch?v=OkxRkNjPHIk}{\textit{Columbia University Campus Tour}}. Other videos are CC licensed (\href{https://www.youtube.com/watch?v=UApe--A1aQM}{\textit{What We Ate Today + Planting Our Garden }}). Part of this research is supported by Mina's Google Ph.D. fellowship. 
\end{acks}

\bibliographystyle{ACM-Reference-Format}
\bibliography{sample-base}
\clearpage
\onecolumn
\appendix


\section{PARTICIPANTS}

\begin{table*}[htbp!]
\small\sffamily\def\arraystretch{1}\setlength{\tabcolsep}{0.8em}
    \centering
    \begin{tabular}{llllll}
        \toprule
       PID  & Gender & Age & Job & Experience & Videos Created \\
       \midrule
        1 & Male & 37 & Video editor & Expert (15 yrs) & Commercials, TV shows, Targetted Ads \\
        2 & Female & 35 & Video editor & Expert (14 yrs) & Commercials, Trailers, Wedding/Birthday videos \\
        3 & Female & 29 &  Video editor & Expert (10 yrs) & Vlogs, Documentaries, Wedding videos \\
        4 & Male & 26 & Studio director & Expert (6 yrs) & Interviews, Training videos, Political Ads \\
        5 & Male & 29 & Video editor & Expert (10 yrs) & Real estate videos, Company Brand/Training videos \\
        6 & Male & 20 & Video editor & Expert (8 yrs) & Documentaries, Social media clips, Sports highlight \\
        7 & Male & 28 & Videographer & Expert (13 yrs) & Music videos, Documentaries, Ads, Short-forms \\
        8 & Male & 32 & Video editor, 2D Animator & Expert (8 yrs) & Promotion videos, Explainer/Tutorial videos, Celebration videos \\
        9 & Male & 22 & Video editor & Expert (4yrs) & Product reviews, Vlogs \\
        10 & Female & 29 & Video editor & Expert (8yrs) & Commercials, Educational videos \\
        11 & Male & 37 & Motion graphics designer & Expert (10yrs) & Corporate videos \\
        12 & Male & 33 & Videographer & Expert (10yrs) & Documentaries, Social media reels \\
        13 & Male & 25 & Motion graphics designer & Expert (6yrs) & Promotional videos, Ads \\
        14 & Male & 35 & Motion graphics designer & Expert (12yrs) & Corporate training videos, Presentations \\
        15 & Male & 18 & Film school student & Amateur (3yrs) & Documentaries \\
        16 & Female & 25 & 3D animator & Amateur (3yrs) & Music videos \\
        17 & Female & 28 & Product manager & Amateur (3yrs) & Personal videos \\
        18 & Male & 24 & Software engineer & Amateur (4yrs) & Personal videos \\
        19 & Male & 33 & Graduate Student & Amateur (6yrs) & Product demo videos \\
        20 & Male & 27 &  Software engineer & Amateur (3yrs) & School projects \\
        21 & Male & 18 & Student & Amateur (4yrs) & Personal videos, Explainers \\
        22 & Male & 43 & Video Editor & Expert (5yrs) & Music/art videos, documentaries \\
        \bottomrule
    \end{tabular}
    \caption{Demographics of study participants (Formative study: P1-P8, Controlled user study: P9-P20, Exploratory case study: P9, P21-P22)} 
    \label{tab:participants}
\end{table*}

\clearpage
\section{Augmentation Prompts}~\label{apndx:augmentation_pipeline}
\begin{table}[h!]
\small\sffamily\def\arraystretch{0.9}\setlength{\tabcolsep}{0.4em}
  \centering
  \begin{tabular}{cccc}
    \hline
    Edit Stage& Attribute & Options \\ \hline
    \multirow{2}{*}{Rough Cut} & Length & 
    \begin{tabular}[c]{@{}l@{}}
    - Make the total duration to be less than 2 minutes. \\ 
    - Make the total duration to be 2-4 minutes. \\ 
    - Make the total duration to be 4-5 minutes.
    \end{tabular} \\ \cline{2-3}
    & Focus & 
    \begin{tabular}[c]{@{}l@{}}
    - Focus on 1-2 sections for a coherent summary, \\ 
    minimize additional sections. \\
    - Focus on 3-4 sections, minimize additional sections. \\ 
    - Extract short parts from many sections for diversity.
    \end{tabular} \\ \hline
    
    \multirow{2}{*}{B-Roll} & Media & 
    \begin{tabular}[c]{@{}l@{}}
    - Illustration \\ 
    - Photo \\ 
    - Video
    \end{tabular} \\ \cline{2-3}
    & Frequency & 
    \begin{tabular}[c]{@{}l@{}}
    - Limit B-Rolls to 2-3. \\ 
    - Limit B-Rolls to 4-5. \\ 
    - Limit B-Rolls to 7-10.
    \end{tabular} \\ \hline
    
    \multirow{2}{*}{Text Effects} & Style & 
    \begin{tabular}[c]{@{}l@{}}
    - Lower \\ 
    - Title \\ 
    - Subtitle
    \end{tabular} \\ \cline{2-3}
    & Frequency & 
    \begin{tabular}[c]{@{}l@{}}
    - Limit pull quote sentences to 2-3. \\ 
    - Limit pull quote sentences to 4-5. \\ 
    - Limit pull quote sentences to 7-10.
    \end{tabular} \\ \hline
  \end{tabular}
  \caption{Augmentation prompts to generate diverse prompts for rough cuts, B-rolls, and text effects. For B-rolls and text effects, the media type is used for searching stock API and embedding text styles using Remotion.}
  \label{tab:video_options}
  \vspace{-3pt}
\end{table}




\section{User Study Examples}
\begin{figure*}[h!]
  \centering
  \includegraphics[width=\textwidth]{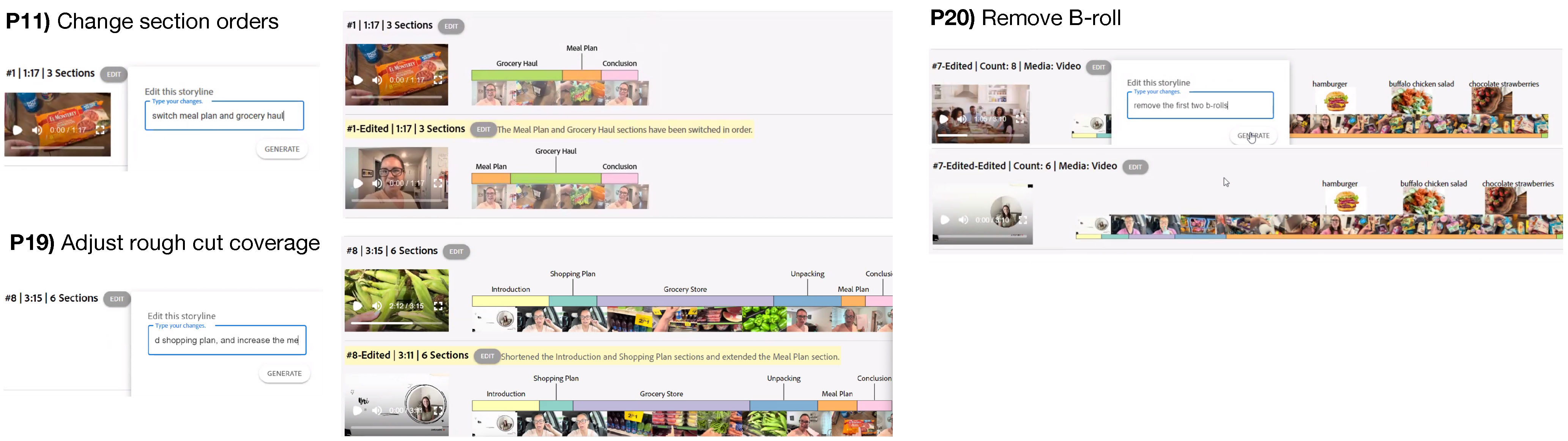}
  \caption{Examples of user edits with VideoDiff}\label{fig:user_edit_examples}
\end{figure*}
\end{document}